# Infinite Drive: Optimal Urban Location of Dynamic Wireless Charging at Signalized Intersections

**Yudai Honma**[a,*], **Daisuke Hasegawa**[b], **Katsuhiro Hata**[c], **Xuesong (Simon) Zhou**[d], **Michael J. Kuby**[e], **Takashi Oguchi**[a]

a) Institute of Industrial Science, The University of Tokyo,
Komaba 4-6-1, Meguro-ku, Tokyo 153-8505, Japan
b) Center for Real Estate Innovation, The University of Tokyo,
Hongo 7-3-1, Bunkyo-ku, Tokyo 113-0033, Japan
c) College of Engineering, Shibaura Institute of Technology,
Toyosu 3-7-5, Koto-ku, Tokyo 135-8548, Japan
d) School of Sustainable Engineering and the Built Environment,
Arizona State University, Tempe, AZ 85281, USA
e) School of Geographical Sciences and Urban Planning,
Arizona State University, Tempe, AZ 85281, USA

* Corresponding author. Email: yudai@iis.u-tokyo.ac.jp

## Abstract

Dynamic Wireless Power Transfer (DWPT) could eliminate plug-in charging in cities, but optimal urban deployment is complex. This paper develops a mixed-integer programming model that optimizes DWPT location under signalized intersection dynamics—acceleration, deceleration, and queue-position-dependent dwell time—through probabilistic signal patterns and saturation headway-based modeling. A case study of Kawagoe City, Japan, shows that electrifying 1.5% of the road network is sufficient to sustain continuous urban EV operation without plug-in charging for the baseline scenario, and at most 2.9% suffices across all tested assumptions. Monte Carlo simulations of continuous trip chains averaging approximately 600 km and reaching up to approximately 800 km confirm that optimized 12 kWh-battery deployments sustain operation in all simulated runs, revealing an infrastructure–battery tradeoff corresponding to roughly 1.7–3.0 tonnes $CO_2e$ of avoided battery manufacturing emissions per vehicle relative to a conventional 40 kWh urban EV. These findings position DWPT deployment as an environmentally efficient pathway for sustainable urban mobility when deployed optimally.

## 1. Introduction

For over a century, automobiles have relied primarily on fossil fuels, contributing significantly to air pollution and climate change. In response, many governments have committed to ending sales of new fossil-fuel-powered cars and vans by 2040 or earlier, and by no later than 2035 in leading markets (UK Government, 2022). As a result, electric vehicles (EVs) are positioned as a central pillar in the decarbonization of transport systems. While EVs eliminate tailpipe emissions, their overall economic and environmental impacts depend critically on the availability of efficient, equitable, and low-carbon charging infrastructure.

Most current EV charging relies on stationary charging at plug-in stations, requiring vehicles to stop for extended periods. This introduces inefficiencies such as route detours, idle queuing, and spatial mismatches between infrastructure and demand. These issues are particularly problematic in dense urban environments, where travel flexibility, energy efficiency, and curbside space are limited. As a result, extensive research has explored the optimal location of charging stations (Coffman et al., 2017; Kchaou-Boujelben, 2021; Rahman et al., 2016), aiming to minimize detours, balance usage, and ensure network-wide accessibility.

To support this, a wide range of facility-location models have been proposed. Classical models such as the *p*-median problem minimize average travel distance to charging sites (ReVelle and Swain, 1970), while the flow-capturing location model (FCLM) identifies station locations that intercept en-route travel flows (Hodgson, 1990). These frameworks have been extended to accommodate path deviations, travel time uncertainty, and stochastic demand (Berman et al., 1995). For EV-specific applications, the flow-refueling location model (FRLM) introduced by Kuby and Lim (2005) explicitly accounts for limited driving range and the need for multiple en-route charging events. More recent mixed-integer programming (MIP) formulations incorporate features such as battery charge tracking, station capacity, and layered coverage (Capar et al., 2013; Kim and Kuby, 2012; Lim and Kuby, 2010; Wang and Lin, 2009). Recent data-driven approaches further leverage trip origin–destination data and urban informatics to estimate charging demand and optimize plug-in station placement (Yi et al., 2022).

These models have also highlighted systemic constraints. As EV adoption grows, charging station congestion, long waiting times, and energy delivery bottlenecks at the system level are becoming critical concerns (Bruglieri et al., 2019; Honma and Toriumi, 2017, 2014; Upchurch et al., 2009). Industry responses—such as ultra-fast charging or oversized onboard batteries—can mitigate range anxiety, but they impose substantial costs. High-power charging stresses local electricity grids, while larger batteries increase vehicle weight, reduce energy efficiency, and intensify reliance on scarce materials such as lithium and cobalt. Frequent high-voltage charging may also accelerate battery degradation, shortening battery life. These challenges underscore the need for infrastructure strategies that address urban EV energy supply in a system-level and resource-efficient manner.

Dynamic Wireless Power Transfer (DWPT) has emerged as a promising alternative. By enabling EVs to receive energy while in motion via inductive coils embedded in roadways, DWPT shifts the charging paradigm from stationary refueling to integrated, en-route energy delivery (Bi et al., 2016; Lukic and Pantic, 2013; Miller et al., 2015). This approach reduces range anxiety and queuing delays while offering broader system benefits. In particular, DWPT allows EVs to operate with smaller onboard batteries, reducing vehicle mass, improving energy efficiency, and lowering material demand (Bi et al., 2019).

Technical progress in DWPT has accelerated in recent years, accompanied by a growing number of real-world demonstration projects. Korea's On-Line Electric Vehicle (OLEV) program provided early operational insights through full-scale deployment on public roads (Hwang et al., 2018; Jang et al., 2015; Ko and Jang, 2013). More recently, Electreon has introduced modular roll-out installation technologies and is conducting pilot projects across Europe (Electreon, 2025). In Japan, DWPT-equipped buses have been deployed as part of demonstration projects for Expo 2025 Osaka, Kansai (Japan Association for the 2025 World Exposition, 2025). Together, these demonstrations indicate that DWPT technology is approaching deployment readiness, shifting the research focus from feasibility toward planning and optimization.

Despite this progress, optimal location planning for DWPT—particularly in intersection-dominated urban environments—remains insufficiently understood. Compared to plug-in charging infrastructure, location modeling for DWPT is still relatively new, though rapidly expanding. A key stream of prior work formulates DWPT deployment as a network problem with multiple routes and congestion effects, often using equilibrium-based approaches. For example, Riemann et al. (2015) proposed a flow-capturing model under stochastic user equilibrium. Chen et al. (2016) proposed an optimization model for deploying DWPT charging lanes in transportation networks under user equilibrium. Manshadi et al. (2018) examined coupled electricity–transportation interactions. Liu et al. (2021) incorporated electricity prices into DWPT location decisions. More recently, Ngo et al. (2020) addressed optimal DWPT positioning across road networks for battery electric vehicles, and Tran et al. (2022) developed an urban DWPT lane location model incorporating dynamic route-choice behavior. Beyond equilibrium formulations, researchers have explored alternative modeling paradigms, including heuristic economic design (Ko et al., 2015), joint planning of stationary and dynamic charging (Chen et al., 2017), and robust optimization approaches (Alwesabi et al., 2022; Liu and Song, 2017).

In parallel, economic feasibility and implementation considerations have been emphasized as essential for real-world viability. Prior studies have examined DWPT deployment at metropolitan and corridor scales, addressing cost effectiveness and system design under realistic spatial constraints (Fuller, 2016; Honma et al., 2024; Mubarak et al., 2021; Trinko et al., 2022; Yan et al., 2022), while other work has focused on pricing and broader economic rationality (He et al., 2013; Jeong et al., 2015). These strands collectively underscore that DWPT planning requires both network-level optimization and careful attention to operational realism.

Most existing DWPT location models, however, are developed for highway corridors or represent urban networks at a coarse resolution, without explicitly capturing the intersection-dominated stop-and-go dynamics that shape charging opportunities in cities. While Zhang et al. (2021) examined eco-driving control for connected and automated electric vehicles at signalized intersections with wireless charging, their focus was on vehicle-side speed optimization rather than citywide DWPT location decisions. In urban contexts, traffic signals, queue formation, and acceleration–deceleration behavior jointly determine (i) the time vehicles spend over embedded coils and (ii) the corresponding energy consumption during approach, stopping, and discharge. DWPT installations near traffic signals may therefore be especially effective, but their benefits are highly sensitive to signal timing, queue-position-dependent dwell times, and approach speeds. Determining where and how much DWPT to install in cities thus requires a high-resolution modeling framework that integrates traffic kinematics, queuing behavior, and energy transfer dynamics.

In this study, we refer to these combined effects of signal timing, queue formation, and acceleration–deceleration behavior near intersections as *signalized intersection dynamics*, which fundamentally govern both EV energy consumption and in-motion charging opportunities in urban networks. **Fig. 1** conceptually illustrates how DWPT effectiveness can differ between highway and urban settings due to these intersection-driven dynamics (Japan Science and Technology Agency, 2025).

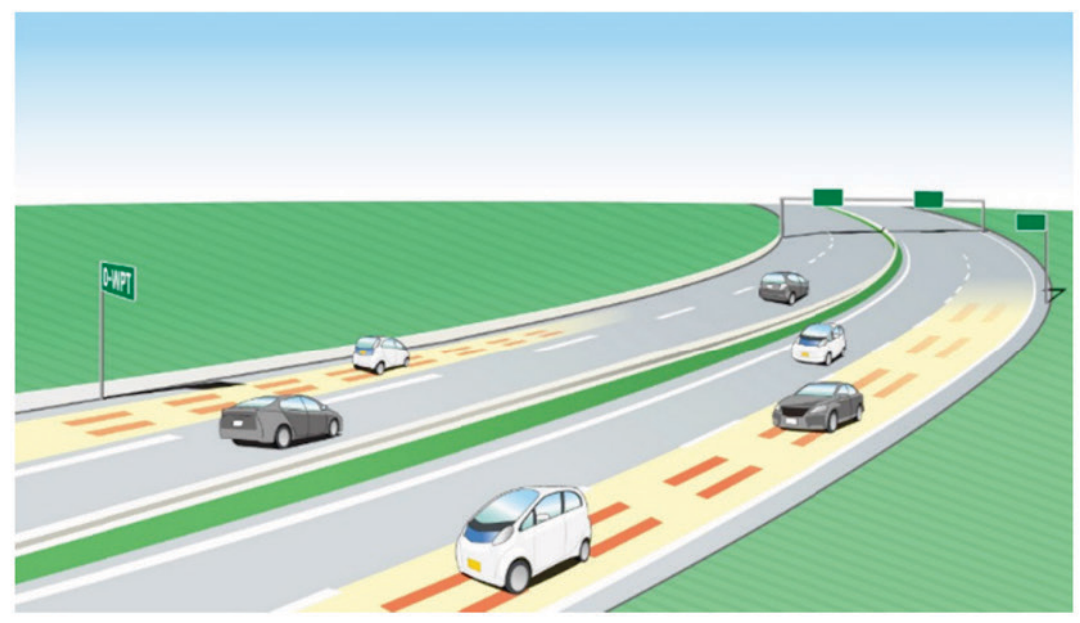

(a) DWPT in expressway scenario

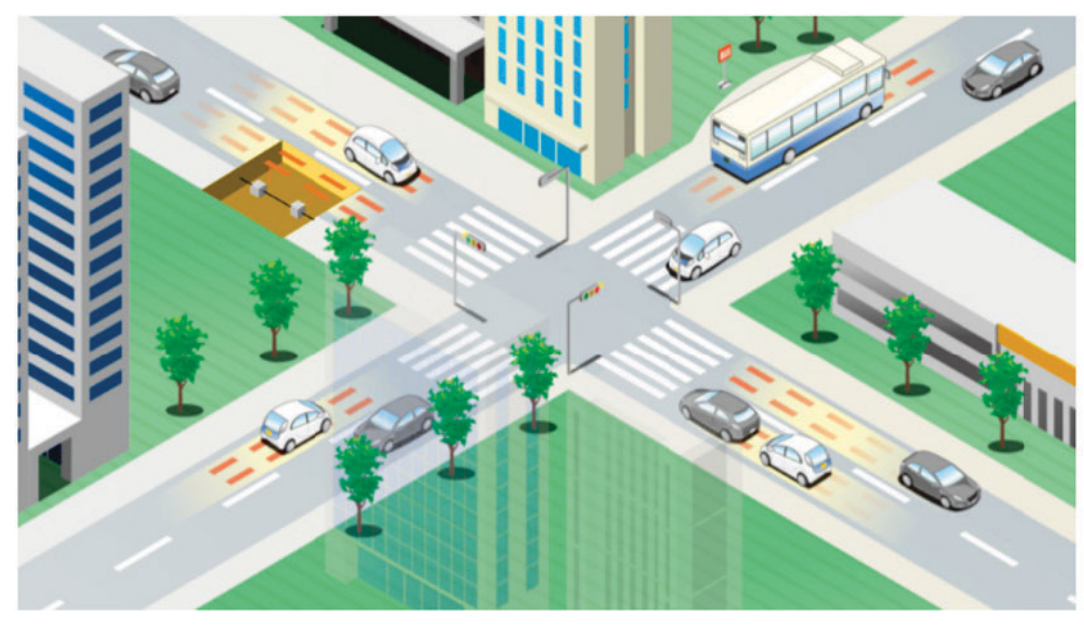

(b) DWPT in urban-scale scenario

**Fig. 1. Graphical image of DWPT location in various scenarios from Japan Science and Technology Agency (2025)**

A critical foundation for DWPT location optimization lies in understanding how vehicle kinematics—particularly acceleration and deceleration near intersections—affect both energy consumption and charging opportunity. **Table 1** summarizes the methodological streams that inform this study. The kinematic equations developed in this study (**Eqs. 13–21**) generate segment-level trajectory data—the shared foundation across all streams.

**Table 1**
**Methodological foundations linking vehicle kinematics, environmental assessment, and charging infrastructure**

| Research Stream | Key Contributions | Relevance To This Study |
|---|---|---|
| **Queuing & traffic flow at intersections** | Lawson et al. (1997): input-output diagrams for queue dynamics at bottlenecks and saturation headway-based modeling | Foundation for dwell time calculation and segment-level trajectory near signals |
| **Cross-resolution traffic-emission integration** | Zhou et al. (2015): mesoscopic simulator with emission model<br>Zhang and Qian (2023): large-scale networks<br>Jiang et al. (2017): urban expressways | Segment-level speed profiles serve as shared inputs for emission estimation and energy transfer calculation |
| **Car-following & emission estimation** | Meng et al. (2021): Newell's model with stochastic parameters for emissions<br>Das and Tanvir (2024): physics-informed LSTM for emission-aware dynamics | Acceleration variability in stop-and-go traffic determines both emission rates and DWPT charging duration |
| **Environmental exposure & accessibility** | Vallamsundar et al. (2016); Zhong et al. (2023): population exposure<br>Song et al. (2017): green accessibility metrics | Spatial distribution of infrastructure determines network-wide environmental outcomes |
| **Eco-routing & emission-sensitive operations** | Pahwa and Jaller (2024): freight eco-routing<br>Köster et al. (2018): emission-sensitive delivery<br>Poulhès and Proulhac (2021): low emission zones | Strategic infrastructure deployment yields benefits beyond immediate locations |
| **Multi-layer optimization & computational methods** | Zhou et al. (2025): flow-through tensor architecture<br>Chen et al. (2024): joint trajectory-signal optimization<br>Wang et al. (2018): DTA review | Unified frameworks for coupling traffic, energy, and infrastructure decisions |

The connection across these streams is direct: the same acceleration and deceleration patterns that drive instantaneous emission rates also determine energy consumption and—for DWPT—the duration vehicles spend over charging coils. Our model builds on this foundation by developing segment-level kinematic equations that capture speed profiles shaped by signal timing and queue discharge—generating trajectory data structures essential for both infrastructure optimization and subsequent environmental assessment.

This study addresses this gap by developing a novel MIP model to optimize the location of DWPT infrastructure for urban EV mobility. Our work makes three primary contributions.

First, we introduce the concept of "Infinite Drive" as a system-level design principle for urban charging infrastructure. Rather than representing a deterministic guarantee at the individual vehicle level, Infinite Drive describes a planning objective under which urban EV operations are structurally energy-sufficient over repeated trip-making cycles, without reliance on *any* plug-in charging. Specifically, we solve for the number, locations, and lengths of DWPT segments that minimize the total installed DWPT length required to sustain such operations. While this goal may seem ambitious now, it is motivated by both medium- and long-term applications. In the medium term, it is especially valuable for drivers who cannot charge at home or work. In fact, DWPT offers the ultimate in convenience for the entire driving population. In the long-term, it leverages the advantage of continuously-on charging and aligns with future scenarios of autonomous-driving societies. In such a future, autonomous vehicles (AVs) could operate continuously, potentially serving logistics needs around the clock, maximizing asset utilization. This makes DWPT particularly valuable, as it allows for continuous charging without the need for AVs to pause their operation.

Second, the proposed modeling framework explicitly incorporates signalized intersection dynamics by accounting for EV acceleration, deceleration, and stopping behavior induced by traffic signal patterns. As vehicles approach signalized intersections, they tend to slow down or stop, increasing the time spent over embedded charging coils and thereby enhancing the amount of energy transferred via DWPT. These signal-induced acceleration–deceleration processes generate critical trade-offs, including whether DWPT should be placed on high-traffic or low-traffic roads and how far upstream from an intersection charging segments should extend. To reflect realistic queuing behavior and stop durations, the model integrates saturation headway-based queuing representations with segment-level kinematic equations. This formulation explicitly captures the interplay between traffic dynamics and charging opportunities, providing a nuanced and operationally realistic representation of DWPT effectiveness in urban networks.

Third, we demonstrate the practical applicability of the proposed framework through a detailed case study of Kawagoe City, a typical medium-sized Japanese city. Using calibrated traffic flow data, we determine the share of the urban road network required to achieve system-level energy sufficiency, examine the sensitivity of this requirement to alternative aggregation assumptions and safety margins, and validate the operational feasibility of the optimized deployment through Monte Carlo simulations of continuous trip chains under finite battery capacity. This analysis further enables explicit characterization of the infrastructure–battery tradeoff and provides a basis for joint decisions on DWPT location and onboard battery sizing. Together, these contributions position targeted DWPT deployment as a strategic component of sustainable urban EV infrastructure planning.

## 2. Conceptual framework

### 2.1. Concept of Infinite Drive

The core objective of this study is to formulate a suitable mathematical optimization model that determines the optimal location of DWPT infrastructure to enable "Infinite Drive"—a scenario in which EVs can meet all urban mobility needs without ever requiring plug-in charging.

Historically, EV infrastructure location models have focused on maximizing demand coverage under resource constraints, as exemplified by the flow-capturing location model (Hodgson, 1990), the flow-refueling location model (Kuby and Lim, 2005), and corridor-scale DWPT planning models (Honma et al., 2024). These models aim to intercept as much travel flow as possible given limited infrastructure budgets. Meanwhile, some station models—starting with Wang and Lin (2009)—take a stricter stance by requiring complete coverage and minimizing the number or cost of stations. These use the location set covering approach, which was first formulated by Toregas et al. (1971) for minimizing the number of facilities required for complete coverage of the urban population by essential emergency services such as fire stations and ambulance depots.

While the Infinite Drive concept is mathematically related to set-covering, it *extends and reinterprets* that logic for a dynamic, long-term application. Traditional set-covering for charging station networks asks, "Can all EV trips be completed?"—a question that is usually trivial in urban settings due to short trip lengths and access to home or workplace charging. In contrast, Infinite Drive asks: *Can all EVs operate indefinitely in urban environments without plugging in at all?* It imposes a perpetual energy sufficiency constraint, ensuring that every likely trip—now and in the future—is energetically feasible solely through en-route charging.

This distinction marks a shift from ensuring the feasibility of individual trips to enabling sustained operational cycles—making the Infinite Drive model notably more ambitious. For everyday urban mobility in all but the largest metropolitan areas, and for early adopters who typically have access to home or workplace charging, driving range is typically not a limiting factor. Infinite Drive becomes more important, however, at the megalopolitan scale, for residents of multifamily housing, and especially for high-utilization vehicles such as taxis, delivery fleets, ride-hailing services, and future autonomous vehicles (AVs) that operate continuously (Axsen and Pickrell-Barr, 2024; Kuby et al., 2025). For these users, Infinite Drive provides complete freedom from plug-in dependence, maximizing operational flexibility, energy access, and vehicle productivity.

In essence, while our model builds on the mathematical structure of the Set Covering Location Problem—locating the fewest facilities to cover all demands—it retools the concept for a dynamic wireless power network and a long-term urban mobility context. DWPT is not just a substitute for plug-in chargers, but a foundation for a fundamentally different energy delivery paradigm.

### 2.2. Critical tradeoff structures

A central consideration in optimizing DWPT location for urban mobility is the effect of vehicle acceleration and deceleration. These kinematic fluctuations significantly influence energy transfer rates and thus the effectiveness of DWPT infrastructure. It is almost intuitive to suggest that "DWPT should be placed near intersections," since vehicles naturally slow down or stop in these areas, increasing dwell time over embedded coils. While this logic is compelling, the optimal location of DWPT is not as straightforward as it may appear. As illustrated in **Fig. 2**, even at intersections, nuanced tradeoffs arise that require rigorous modeling.

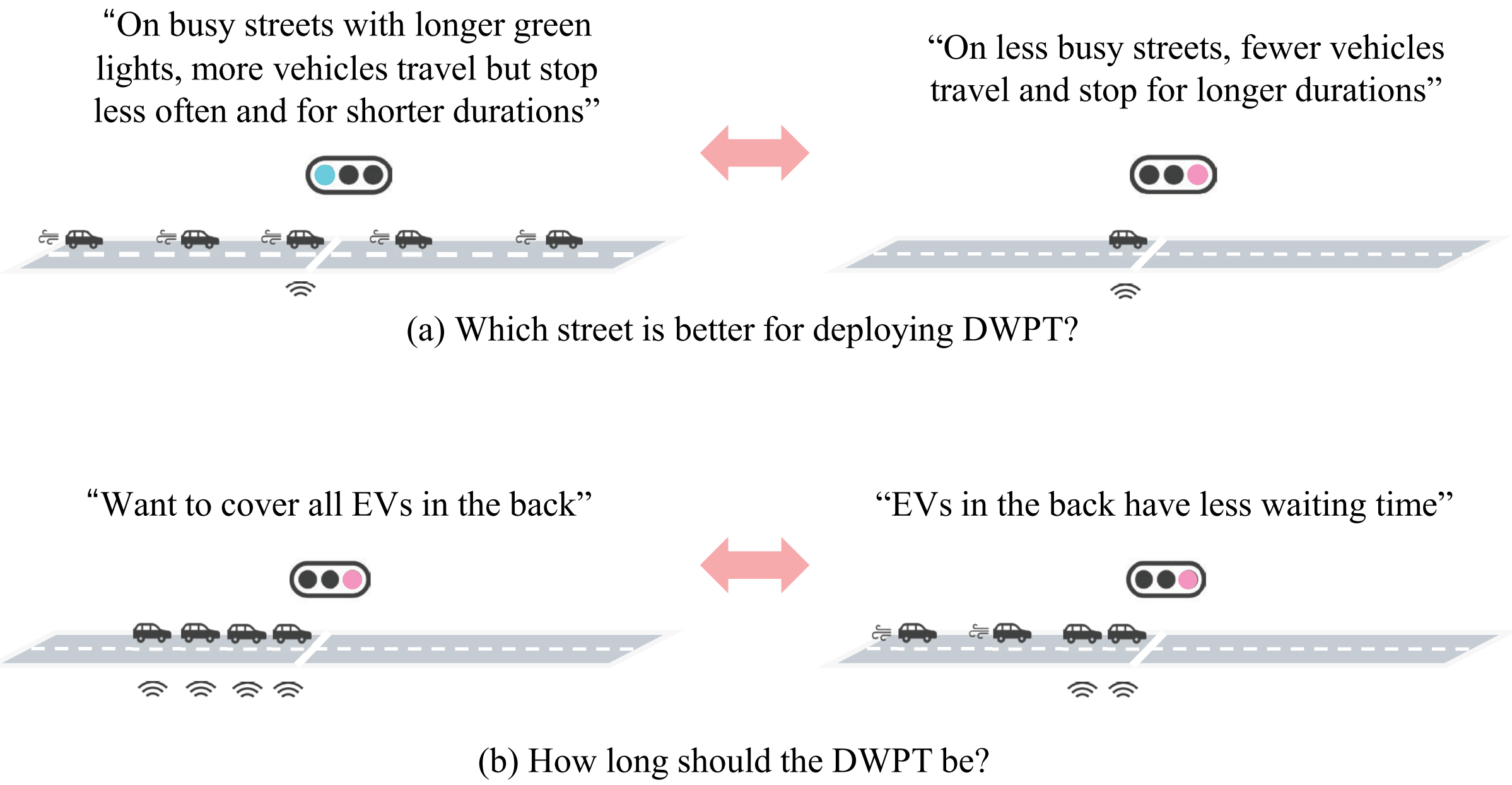

**Fig. 2. Non-obvious tradeoffs at intersections**

The first tradeoff (**Fig. 2(a)**) concerns *which streets should be prioritized for DWPT installation*. Roads with high traffic volumes appear attractive due to the large number of vehicles they serve, but these roads typically have longer green-light durations to manage throughput, meaning vehicles stop less frequently and for shorter durations. This reduces the opportunities for efficient charging. On the other hand, lower-volume roads may feature longer red lights, allowing for longer stop times and potentially more effective energy transfer—but these roads see fewer vehicles overall. Thus, there is a tension between maximizing the number of charging opportunities and the quality of each opportunity.

The second tradeoff (**Fig. 2(b)**) involves *how long the DWPT segment at a given intersection should be*. Even though the common terminology for this technology is "*Dynamic*" WPT, it is most efficient to provide substantial power transfer when vehicles are stopped at red lights or at least slowing down for turns. Therefore, placing DWPT starting from the stop lines at traffic signals is ideal. However, determining how far back these systems should extend is challenging because the queue of EVs waiting at a red light gradually lengthens. Covering a longer segment would allow more EVs to benefit from the DWPT, but the queue will less frequently extend back that far and when it does the EVs will have shorter stop durations.

Together, these tradeoffs underscore the importance of detailed, segment-level modeling to determine where and how DWPT should be deployed. Any effective optimization framework must account for signal timing, traffic volumes, queuing behavior, and the kinematic states of vehicles to ensure that installations maximize real-world utility and efficiency, as shown in **Fig. 3**, where s*egment-level trajectory data serve as the shared foundation across all three domains.*

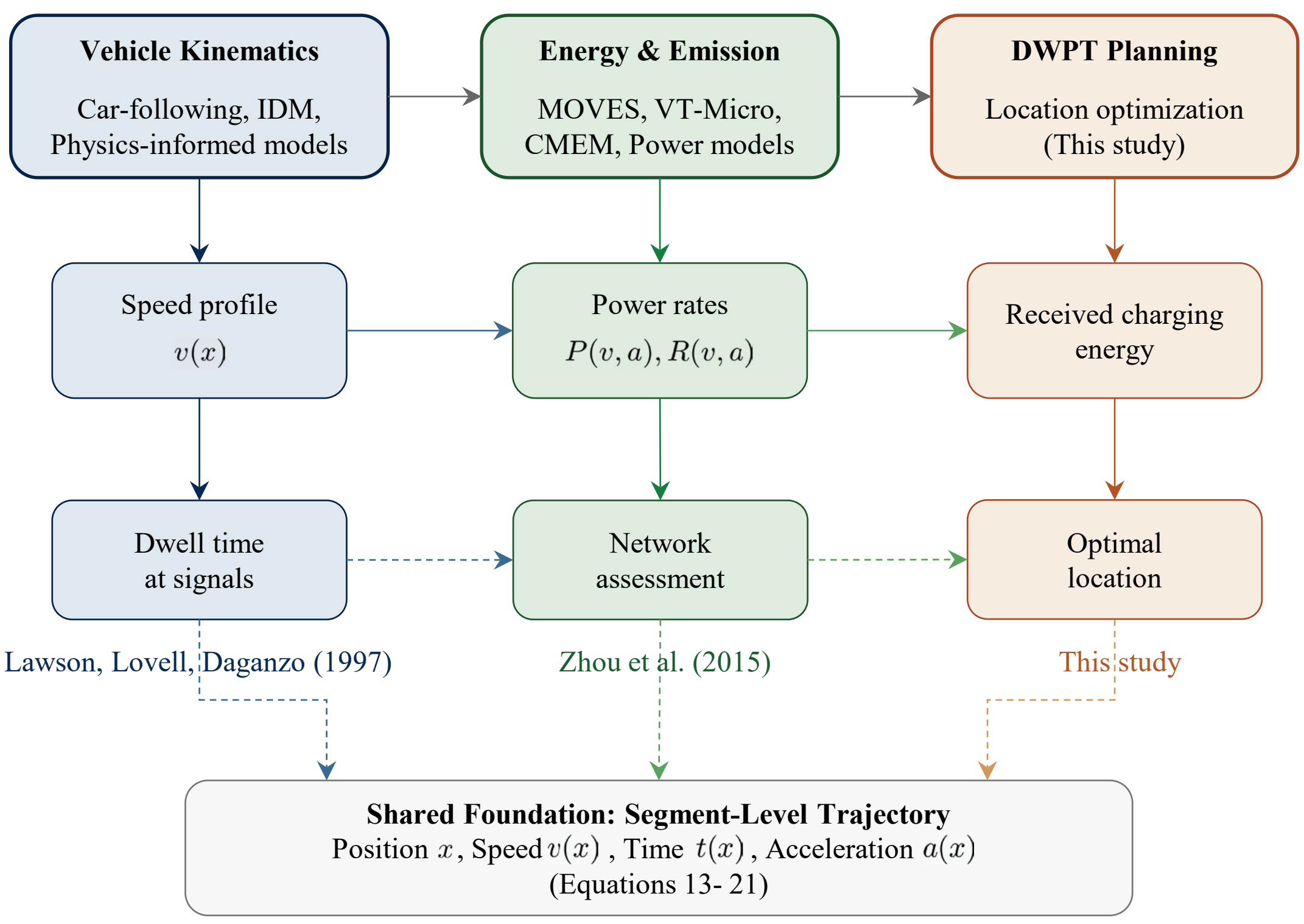

**Fig. 3. Conceptual framework linking car-following kinematics, energy/emission estimation, and DWPT infrastructure optimization.**

## 3. Formulation

### 3.1. Model assumptions

The model is built on several key assumptions designed to balance realism and computational tractability. First, the road network, EV routes, and associated traffic volumes are treated as exogenous inputs. Each OD pair may have multiple pre-defined routes, but route choice behavior is fixed and does not respond to DWPT locations. This simplification not only reflects a scenario in which DWPT infrastructure is seamlessly integrated into everyday traffic patterns but also avoids the computational complexity of equilibrium traffic assignment while still capturing realistic route-level variations in flow volumes. The assumption is most appropriate when DWPT is sparsely deployed and not associated with explicit pricing incentives, so that the detour effect on route choice remains small; as DWPT density or pricing increases, equilibrium-based formulations such as those of Riemann et al. (2015) and Chen et al. (2016) provide the natural complementary perspective. The present framework thus occupies the fixed-flow planning end of a methodological spectrum, prioritizing spatial resolution of signalized-intersection charging dynamics over equilibrium feedback effects.

Second, while signal-level vehicle trajectories are indeed critical to determining the optimal locations of DWPTs as noted earlier, the model accounts for these variations through probabilistic signal patterns. Because it is practically infeasible to collect highly granular data—such as the precise stopping position of each individual vehicle and the corresponding signal state at that moment—the model instead employs signal probability distributions to represent the likelihood of EVs stopping at traffic signals. For each intersection, a discrete set of patterns is used to capture the probabilities of stopping or passing through during a given signal cycle. These patterns encode the likelihood of an EV occupying a specific position in the queue, which directly affects its dwell time and, consequently, the amount of energy transferred via the DWPT.

Third, to reflect spatial granularity, the network is discretized into 7 m segments, approximately corresponding to the space occupied by a single stopped vehicle at a red light. This segmentation enables the model to explicitly capture acceleration, deceleration, and queue-position effects at intersections. It also allows DWPT location decisions to be expressed at the segment level.

Finally, the model's objective is to minimize the total installed length of DWPT infrastructure required to achieve "Infinite Drive" within an urban network, subject to constraints on energy-sufficiency. Theoretically, these constraints could be imposed on each origin–destination route, but that would not only be computationally infeasible for larger applications but would also lead to substantial overbuilding of charging capacity, as we explain later. For now, suffice it to note here that the model aggregates routes into "energy balancing groups," with a constraint requiring that the aggregate energy balance remains non-negative for each group at all times.

Relative to existing corridor-scale DWPT location models (Honma et al., 2024), the proposed framework introduces four key methodological extensions tailored to urban settings: (i) probabilistic signal patterns that capture stochastic stop-and-go behavior at signalized intersections; (ii) saturation headway-based queue modeling that links queue position to dwell time and charging duration; (iii) segment-level kinematic equations that explicitly model acceleration and deceleration around intersections; and (iv) operational validation through state-of-charge simulations with finite-capacity batteries (Section 6). Together, these extensions enable Infinite Drive analysis at fine spatial granularity in signal-controlled urban networks, where signal-induced variability dominates energy transfer opportunities.

The framework also positions itself within a broader methodological landscape for coupling traffic dynamics with energy costs. Alternative approaches, such as Eco-System Optimal (ECO-SO) flow assignment (Lu et al., 2016), recent dynamic user equilibrium formulations with non-monotonic emission costs (Tidswell et al., 2021), and the hyper-prism accessibility structure of Mahmoudi et al. (2019), endogenize driver responses to infrastructure conditions but become computationally prohibitive at the segment-level resolution required to capture signalized-intersection charging dynamics. The present framework therefore trades behavioral feedback for spatial fidelity, accepting fixed route flows in exchange for explicit segment-level modeling of signal-induced acceleration, deceleration, and queue-position-dependent dwell time.

### 3.2. Mathematical optimization problem

All notations for the model are defined below:

Indices:

| | |
|---|---|
| $i$ | Index of small (7 m) segments that divide road networks |
| $s$ | Index of signal patterns, including vehicle stop positions at red lights |
| $r$ | Index of routes |
| $k$ | Index of energy balancing groups (mutually exclusive route subsets) |

Sets:

| | |
|---|---|
| $I$ | Set of small segments that divide road networks |
| $I_r$ | Set of small segments that divide road networks for route $r$ |
| $S$ | Set of signal patterns, including vehicle stop positions at red lights |
| $R$ | Set of all routes |
| $R_k$ | Set of routes in energy balancing group $k$ |
| $K$ | Set of energy balancing groups |

Parameters:

| | |
|---|---|
| $b_r$ | Expected energy balance for route $r$ (kWh) |
| $c_{is}^r$ | Required electric energy for route $r$ to pass through segment $i$ given signal pattern $s$ (kWh) |
| $d_i$ | Length of segment $i$ (m) |
| $f_r$ | Flow volume of route $r$ |
| $g_{is}^r$ | Electric energy transfer for route $r$ from DWPT at segment $i$ given signal pattern $s$ (kWh) |
| $p_{is}^r$ | Probability of encountering signal pattern $s$ at segment $i$ (dimensionless) |
| $\varphi$ | Safety coefficient for energy balance |

Decision Variables:

| | |
|---|---|
| $z_i$ | 1 if DWPT is installed on segment $i$, 0 otherwise |

As outlined above, the model minimizes the length of DWPT installation necessary to enable all EVs to complete all their routes to achieve Infinite Drive within an urban area. Note that there is only one type of decision variable, $z_i$, representing whether a 7-m DWPT coil is embedded in the roadway on segment *i* or not. The second type of binary decision variable typically found in charging station location models based on the concept of max-covering—representing whether an OD pair is covered or not—is not needed in this formulation because all routes must be covered. In addition, no length variable is needed because the roads are divided into individual segments *i.* No route-used variable is needed because routes are given. Finally, no variables are included to explicitly keep track of the battery state-of-charge (SOC) on each segment of each route, because the model imposes an expected-value energy-balance constraint at the route and group level rather than tracking deterministic SOC trajectories—a formulation consistent with the probabilistic signal-pattern representation. Finite-battery operational feasibility is validated separately through the SOC simulations in Section 6.

The mathematical optimization problem proposed is as follows:

$$\text{Min.} \quad \sum_{i\in I} d_i z_i \tag{1}$$

subject to:

$$\sum_{i\in I_r} \sum_{s\in S} p_{is}^r (g_{is}^r z_i - \varphi\, c_{is}^r) = b_r \qquad \forall r \in R \tag{2}$$

$$\sum_{r\in R_k} f_r b_r \geq 0 \qquad \forall k \in K \tag{3}$$

$$z_i \in \{0,1\} \qquad \forall i \in I \tag{4}$$

**Eq. 1** minimizes the total length of DWPT installation across the target area. Constraint **Eq. 2** represents the relationship for calculating the energy balance during EV travel along a specific route $r$ in the designated area. In this study, the signal patterns are assumed to be given probabilistically. The term $p_{is}^{r}$ accounts for the probability of each signal pattern, including the specific position where an EV stops at a red light. Changes in signal patterns and the stopping position affect the timing of acceleration and deceleration, which in turn influences the amount of energy $g_{is}^{r}$ that can be transferred from the DWPT and the electric energy consumption $c_{is}^{r}$ of the EV (obtained by integrating instantaneous power over time). To embed a safety margin directly into the energy balance, the consumption term is scaled by a safety coefficient φ $(> 1)$, so that the pattern-specific balance on each segment is given by the transferable DWPT energy $g_{is}^{r} z_i$ minus the scaled consumption φ $c_{is}^{r}$; the role of φ is discussed below. Consequently, the expected energy balance can be calculated by multiplying the probability of each signal pattern $s$ by the corresponding energy balance for that pattern and then summing these products across all road segments $i$ used by route $r$.

Constraint **Eq. 3** addresses the energy balance condition necessary for achieving Infinite Drive, introducing two key concepts. The first concept is the "energy balancing group" $k$, which groups several routes $r \in R_k$. While it is possible to impose a constraint ensuring a positive energy balance for each individual route $r$, this approach is clearly an overestimate. In reality, EVs traverse various routes among different origins and destinations, and it suffices to ensure a positive overall energy balance as a whole. Thus, on the left-hand side, we calculate the expected total energy balance for each group $k$ and constrain it to be non-negative. Flows $f_r$ are used as weights to represent the relative frequency of route usage within a group and can be interpreted as probability weights for a representative vehicle's repeated trip-making cycle. The aggregation does not imply physical energy sharing across vehicles; rather, it captures expected energy balance over time for an individual vehicle following typical urban travel patterns. The concept of the groups will be discussed in more detail in the next sub-section. The second key concept is the safety coefficient $\varphi$, which enters the route-level energy balance in **Eq. 2** by scaling the consumption term. Requiring the expected DWPT energy gain to exceed φ times the expected consumption—rather than merely to match it—imposes a deterministic safety margin (for example, $\varphi = 1.1$ requires a 10% energy surplus), which is precisely why a larger φ calls for more DWPT infrastructure to satisfy the constraint. While the formulation does not explicitly track battery state-of-charge dynamics, the safety coefficient $\varphi$ provides a deterministic-equivalent of a chance constraint, imposing a tolerance margin on the group-level expected-energy-balance guarantee. This formulation accommodates several practical sources of variability that are difficult to capture explicitly within the optimization, including heterogeneity in onboard battery capacity across the EV fleet by model and vintage, variability in DWPT coil spacing and energy transfer efficiency, road grade effects (hills) not captured by the segment-averaged kinematic equations, and ambient temperature effects on motor and battery performance. By choosing $\varphi > 1$ as a standard engineering convention, planners introduce a deterministic safety margin against these stochastic and fleet-level uncertainties without expanding the dimensionality of the optimization problem. Finally, Constraint **Eq. 4** requires the variables related to the installation of DWPT to be binary.

This mathematical optimization problem, while appearing straightforward, is not easily solvable due to several complexities. Firstly, the probabilistic patterns related to signals $p_{is}^{r}$, the amount of energy that can be transferred from DWPT $g_{is}^{r}$, and the energy consumption by EVs $c_{is}^{r}$ must be accurately prepared. Furthermore, the problem involves partitioning the entire road network in the target area—comprising a total length of approximately 150 km, as considered in our numerical example—into 7-m segments, each corresponding to the space occupied by a single vehicle. This segmentation leads to over 21,000 binary variables $z_i$, making the optimization problem computationally demanding.

### 3.3. Definition of "Energy Balancing Groups"

This section provides further explanation regarding the aggregation of selected routes into "energy balancing groups," for which the expected energy balance must be non-negative. While it is theoretically possible to constrain each individual route $r$ to have a positive expected energy balance, this approach would clearly be an overestimate of DWPT needed, and thus economically inefficient. For instance, if the city has numerous very short trips, such as visits to a nearby convenience store, enforcing a positive energy balance for each trip would necessitate the installation of DWPT almost everywhere in the city. This requirement is impractical, given that EVs are equipped with batteries of sufficient capacity to manage such short trips without constant recharging from DWPT.

Instead, we propose using a more natural aggregation scheme: routes aggregated by each location assumed to be an EV driver's home (**Fig. 4(a)**). This means collecting all routes that either start or end at any home locations, including multifamily dwellings. The intention behind this approach is to consider a scenario where drivers make multiple trips to various destinations and return home, ensuring that the overall expected energy balance is met across these trips. By achieving a positive energy balance for the group as a whole, we ensure that the concept of Infinite Drive is maintained. The specific spatial granularity of the "home location" can be tailored to the resolution of the available demand data—possible aggregation levels include individual buildings or households, road links serving as virtual origins, traffic analysis zones (TAZs), or census-defined neighborhoods. Finer aggregation tends to require more extensive DWPT deployment, while coarser aggregation permits greater energy balancing across routes (cf. Section 5.3). This approach allows for a more practical and realistic deployment of DWPT, focusing on aggregate rather than individual route-level energy balances.

Of course, alternative aggregation schemes can also be considered. For example, many optimal EV infrastructure location models allocate demand based on Origin-Destination pairs. Correspondingly, one could aggregate routes based on these OD pairs, encompassing various paths between the same two locations (**Fig. 4(b)**). This approach would be a finer aggregation than the location-based approach, likely requiring more extensive DWPT installations to meet the energy balance criteria.

Conversely, one could consider an even coarser aggregation, where it is sufficient for the entire target area's expected energy balance to be non-negative. This broader scope of aggregation applies when individual specific points or OD pairs do not need to maintain an energy balance independently. Instead, the overall expected energy balance across the entire region would suffice (**Fig. 4(c)**). This scenario represents a pattern that would require the least amount of DWPT installation, assuming an integrated approach to shared vehicle usage that optimizes energy distribution and minimizes infrastructure needs.

These three aggregation schemes naturally correspond to different vehicle classes encountered in real-world fleet operations. Anchor-location aggregation (**Fig. 4(a)**) is well-suited to privately owned EVs whose trips anchor at home, as well as to freight and delivery vehicles whose trip chains begin and end at centralized depots. OD-pair aggregation (**Fig. 4(b)**) suits operations dominated by specific recurring trip patterns. Region-based aggregation (**Fig. 4(c)**) is most appropriate for shared mobility services and future autonomous vehicles whose operations may span the network without a fixed anchor. Because the energy-balancing group set $K$ is defined as a generic partition of the route set $R$, the proposed framework accommodates all such vehicle classes—and arbitrary mixtures thereof—without modifying the underlying optimization formulation.

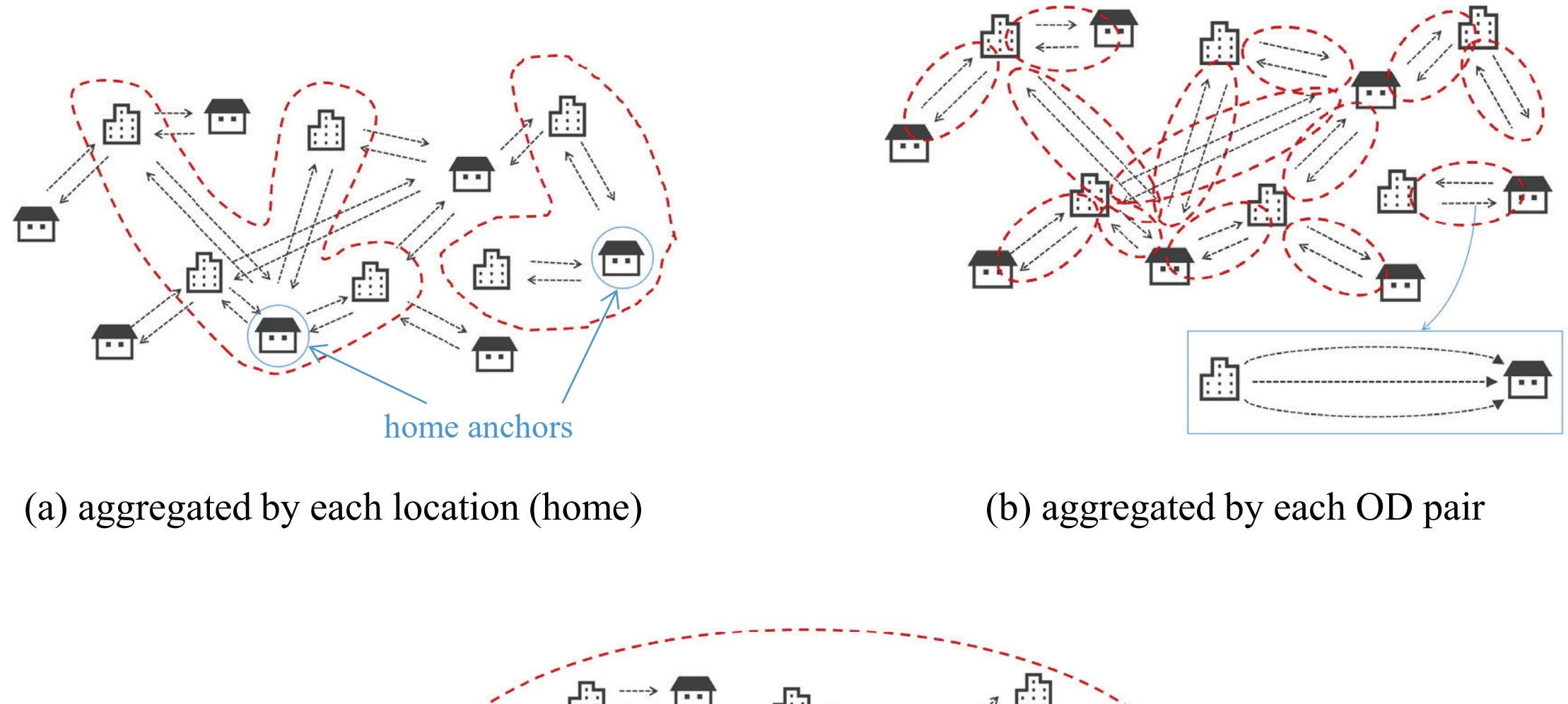

(a) aggregated by each location (home)

(b) aggregated by each OD pair

(c) aggregated across the entire region

**Fig. 4. Three approaches to grouping routes into "energy balancing groups."**

### 3.4. Considerations of signal patterns *s*

Next, we consider the probability distribution of signal patterns, because capturing or simulating stops at traffic signals and the precise timing of acceleration and braking is challenging. Therefore, we supplement the exogenous routing data through probabilistic calculations for detailed aspects such as red-light cycles and the vehicle position in the queue.

In this study, the discrete index $s$ encodes the joint state of (i) whether the signal is green or red, and (ii) the queue position of the vehicle if the signal is red. Specifically, $s \in \{0, 1, 2, \dots, L_{\mathrm{red}}\}$, where $s = 0$ denotes a green-light passage, and $s = 1, 2, \dots, L_{\mathrm{red}}$ denotes a red-light stop at the $s$-th position in the queue. This single-variable encoding captures both the signal phase and the queue dynamics in a unified formulation. The probabilities of encountering green and red lights can be expressed as follows:

Probability of red light: $$q_{\mathrm{red}} \tag{5}$$
Probability of green light: $$1 - q_{\mathrm{red}} \tag{6}$$

Once the probabilities of signal patterns are determined, we estimate how many vehicles queue up during a red-light cycle:

Time of red light: $$W_{\mathrm{red}} = c \times q_{\mathrm{red}} \tag{7}$$
Queue length during red light: $$L_{\mathrm{red}} = \frac{c}{3600} \times q_{\mathrm{red}} \times f \tag{8}$$

Here, $f$ represents the number of EVs per hour per link, and $c$ is the cycle time, calibrated per intersection from observed signal data. For roads with multiple lanes, the value of $f$ is divided by the number of lanes.

To more accurately reflect queuing behavior, we incorporate the concept of saturation headway (Bonneson, 1992; Kimber et al., 1986), the average time interval between successive vehicles as they depart once the signal turns green. The probability of being the $s$-th vehicle in the queue and the corresponding waiting time are given by:

$$\text{Probability of being the } s\text{-th vehicle:} \quad \frac{q_{\text{red}}}{L_{\text{red}}} \tag{9}$$

$$\text{Waiting time of being the } s\text{-th vehicle:} \quad \frac{W_{\text{red}}}{L_{\text{red}}} \times (L_{\text{red}} - s + 1) + h_s \times (s - 1) \tag{10}$$

where $h_s$ is the saturation headway (typically 2.0–2.5 seconds). This formulation better captures real-world departure delays, particularly for vehicles farther back in the queue. This uniform approximation reflects the assumption that vehicle arrival times are evenly distributed within the signal cycle, which is a common and tractable representation when detailed arrival profiles are unavailable. In coordinated-signal corridors where platoon arrivals predominate, $p_{is}^{r}$ would deviate from uniform, and the formulation can be readily recalibrated with the corresponding non-uniform arrival profile to preserve transferability to such settings.

Given that the domain of $s \in \{0, 1, 2, \ldots, L_{\text{red}}\}$, the probability distribution $p_{is}^{r}$ for signal pattern $s$ is defined as:

$$p_{is}^{r} = \begin{cases} 1 - q_{\text{red}} & \text{if } s = 0 \quad \text{(Green light)} \\ \frac{q_{\text{red}}}{L_{\text{red}}} & \text{if } s = 1, 2, \ldots, L_{\text{red}} \quad \text{(Red light)} \end{cases} \tag{11}$$

Additionally, the specific waiting time $t_{is}^{r,\text{red}}$ for stopping at a traffic signal on segment $i$ during a red light is:

$$t_{is}^{r,\text{red}} = \begin{cases} \frac{W_{\text{red}}}{L_{\text{red}}} \times (L_{\text{red}} - s + 1) + h_s \times (s - 1) & \text{if the stop occurs on segment } i \\ 0 & \text{otherwise} \end{cases} \tag{12}$$

This model does not yet account for dedicated lanes for right or left turns. Consequently, the values of $p_{is}^{r}$ and $t_{is}^{r,\text{red}}$ do not depend on route $r$. For future modeling work, however, where such dependencies may be considered, the notation $r$ has been retained in the formulation for generality. Additionally, while the notation does not explicitly include indices for specific intersections or times of day, these factors were individually set per intersection and time slot in this study.

### 3.5. Considerations of acceleration and braking

The specific behaviors of acceleration and braking vary based on whether a vehicle stops at a traffic signal and, if so, its position in the queue. How these variations influence both energy consumption and the amount of time an EV is positioned over DWPT depends on two key elements. The first element is the signal pattern and the associated stopping position. If the light is green, vehicles can pass through the intersection without stopping (**Fig. 5(a)**). However, if the light is red, vehicles must stop at an appropriate position before the intersection and then accelerate again when the light changes (**Fig. 5(b)**).

The second element concerns the vehicle's maneuver at the intersection—whether it goes straight or turns left or right. Since the routes are exogenously determined in this study, we can specify the maneuvers for each vehicle. While vehicles going straight can generally continue without stopping (assuming they have a green light), those making left or right turns must slow significantly or completely stop at the intersection, even if they have a green light, as a safety precaution (**Fig. 5(c,d)**). The conservative assumption to make here is that a full stop is necessary at each turn and the stop is assumed to occur in the central part of the intersection where DWPT is not present. Thus the stop time does not contribute to energy transfer calculations. For vehicles proceeding straight, normal acceleration and braking are considered based on the speed limits of each link.

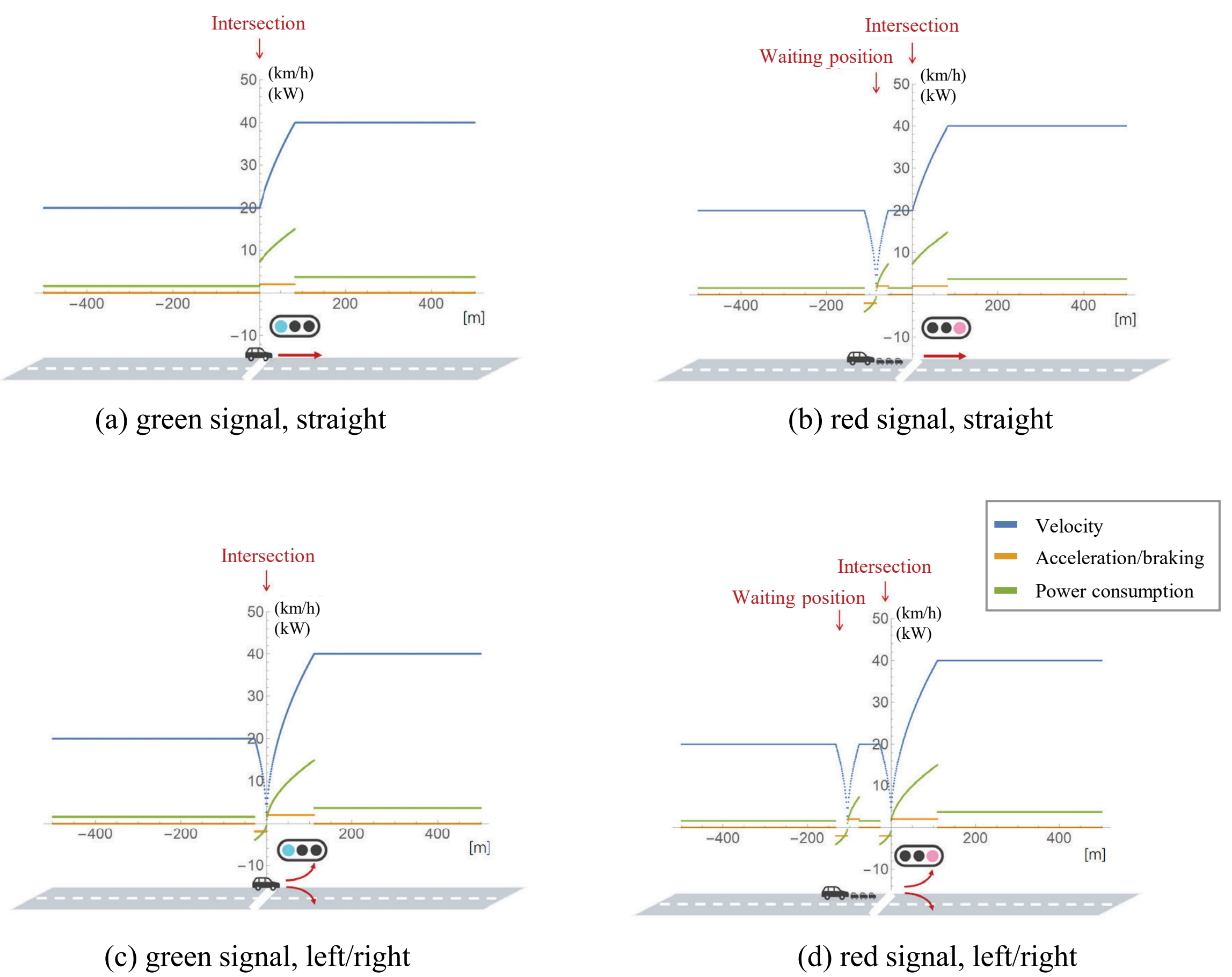

**Fig. 5. Mathematical scenario to reflect signal pattern and vehicle's maneuver**

It is crucial to note that these specific acceleration and braking behaviors are dictated by the probability distribution of signal pattern $s$, as these probabilities will govern the vehicle's interactions with the signal and, consequently, their energy consumption and the duration over the DWPT.

The following equations determine the position, speed, and transit time on each segment due to acceleration and braking. Starting from position $x = 0$ and velocity $v = 0$ with constant acceleration $a$, the relationships between position $x$, velocity $v$, and time $t$ are given by:

[When time $t$ is given:]

$$v(t) = at \tag{13}$$

$$x(t) = \frac{1}{2}at^2 \tag{14}$$

[When velocity $v$ is given:]

$$t(v) = \frac{v}{a} \tag{15}$$

$$x(v) = \frac{v^2}{2a} \tag{16}$$

[When position $x$ is given:]

$$t(x) = \sqrt{\frac{2x}{a}} \tag{17}$$

$$v(x) = \sqrt{2ax} \tag{18}$$

These equations apply to both acceleration and deceleration scenarios. For example, if a vehicle is traveling at a constant speed $v_{const}$ and needs to stop at position $x_{\text{stop}}$, we can calculate the required braking distance using **Eq. 16** as $x_{\text{stop}} - x(v_{const})$. The velocity reduction from $x_{\text{stop}} - x(v_{const})$ to $x_{\text{stop}}$ can then be determined using **Eq. 18** for $v(x)$. Similarly, if a vehicle decelerates from a constant speed $v_{const}^1$ to $v_{const}^2$ at position $x_2$, the transition can be calculated as $x_2 - x(v_{const}^1) + x(v_{const}^2)$.

The critical component, the transit time $t_{is}^{r,\text{move}}$ over each segment $i$, with segment boundaries $[x_i^1, x_i^2]$, can be obtained by integrating the reciprocal of the speed over the segment:

$$t_{is}^{r,\text{move}} = \int_{x_i^1}^{x_i^2} \frac{1}{v(x|s)} \mathrm{d}x \tag{19}$$

Combining the transit time during movement $t_i^{r,\text{move}}(s)$ with the waiting time at red lights $t_{is}^{r,\text{red}}$ from **Eq. 12**, we get:

$$t_{is}^r = t_{is}^{r,\text{move}} + t_{is}^{r,\text{red}} \tag{20}$$

Finally, let $g_{\text{trans}}$ denote the effective DWPT transfer power (kW) (set to 18.7 kW in this study; see Section 4.1). Then the energy received from DWPT during the segment can be calculated as:

$$g_{is}^r = g_{\text{trans}} * t_{is}^r \tag{21}$$

Note that the transit time is converted from seconds to hours when computing energy in kWh. These equations collectively provide the basis for determining the dynamic power transfer and consumption behaviors of EVs as they move through segments with varying speeds and stopping patterns.

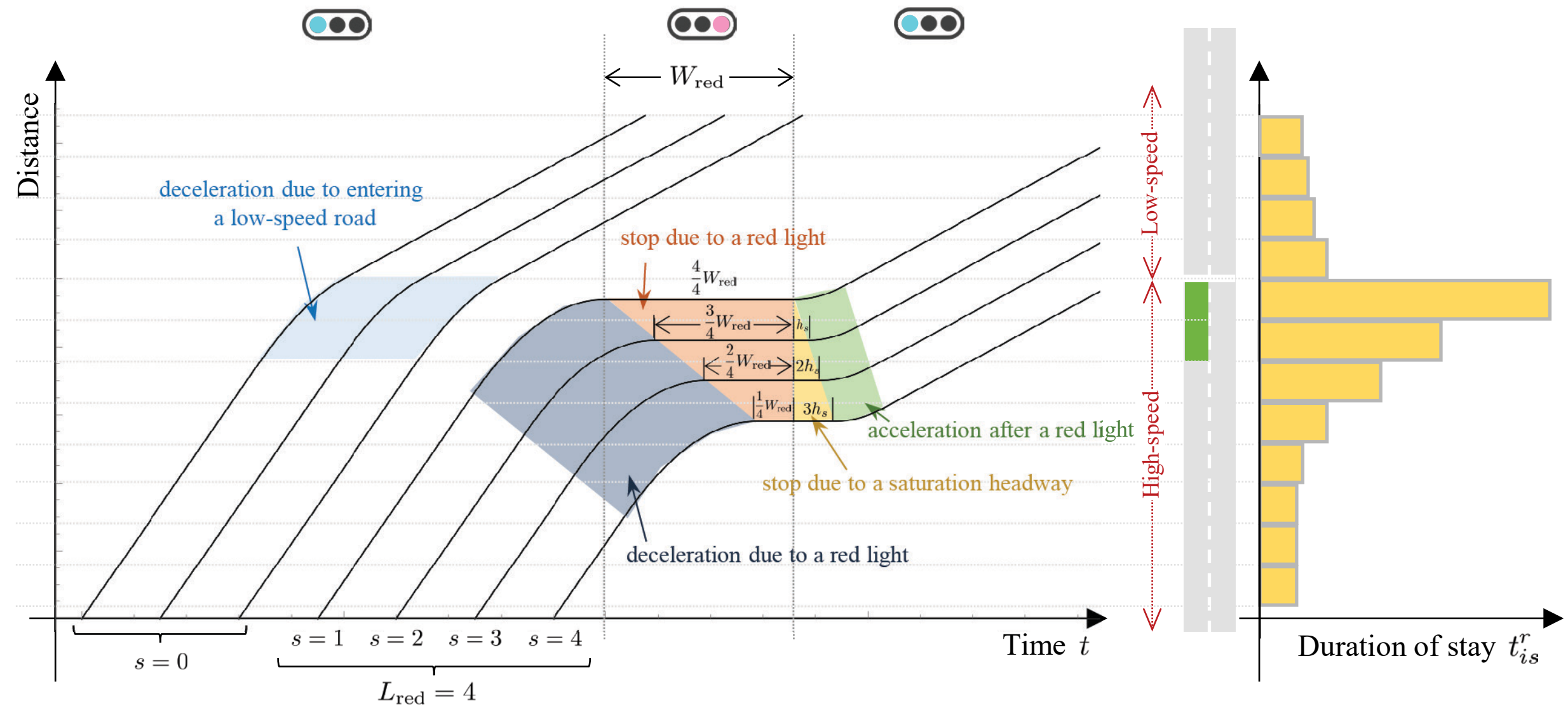

**Fig. 6. Space–time diagram of vehicle trajectories near intersections**

To supplement our modeling framework, **Fig. 6** presents a space–time diagram of vehicle trajectories near intersections. This figure visually demonstrates that dwell times depend on queue position, thereby supporting the formulation used in this study. It also illustrates the deceleration and acceleration phases that arise through interactions with traffic signals. The pink-shaded area represents stoppage due to red lights, while the yellow-shaded area indicates travel on low-speed segments or deceleration when approaching a red light. The green-shaded area reflects acceleration after the signal turns green.

By depicting these speed fluctuations and differences in stop durations across signal phases, the diagram shows that the vehicle dynamics considered in this study are not exogenously imposed but are endogenously determined by signal control, queue position, and intersection behavior. Moreover, by embedding speed variation into segment-level kinematic equations and signal-pattern probabilities, the figure helps address potential concerns regarding the realism of our velocity modeling. Although our model does not implement a fully time-dependent traffic assignment, the deterministic segment-level kinematics—reflecting the impacts of signals and turning movements—appropriately capture the speed variations necessary for this study. This approach offers an effective balance between computational efficiency and behavioral realism for urban-scale DWPT planning.

## 4. Data

### 4.1. Power consumption equation

First, **Eq. 22** computes the power consumption in kilowatts (kW), based on the previous studies (Fiori et al., 2016; The Engineering ToolBox, 2004; Wu et al., 2015):

$$P(v,a,\theta) = \frac{1}{\eta} v \left( ma + mg\cos\theta\, f_{rl} + \frac{1}{2}\rho A_f C_D v^2 + mg\sin\theta \right) \qquad (22)$$

$v$ represents the EV's speed in meters per second (m/s), $a$ the acceleration in meters per second squared (m/s$^2$), and $\theta$ the road gradient in degrees (°) **(Table 2)**. Regarding the effective DWPT transfer power $g_{\text{trans}}$ from DWPT, existing studies suggest an output of 20–25 kW (Hata et al., 2019). In addition, when a vehicle is stationary, a wireless power transfer, such as that of WPT4, has been proposed (SAE J2954 standard, 2020). We assume a transfer capacity of 22 kW with an efficiency of 85%, resulting in an effective output of 18.7 kW.

**Table 2**
**Parameters for calculating motor power**

| Parameters | Values |
|---|---|
| Efficiency of electric motor $\eta$ (%) | 90 |
| Vehicle weight (including driver) $m$ (kg) | 1,640 |
| Gravitational acceleration $g$ (m/s$^2$) | 9.8066 |
| Rolling resistance coefficient $f_{rl}$ | 0.015 |
| Air mass density $\rho$ (kg/m$^3$) | 1.2256 |
| Frontal area of vehicle $A_f$ (m$^2$) | 2.34 |
| Aerodynamic drag coefficient $C_D$ | 0.32 |

### 4.2. Traffic flow data for Kawagoe City

The case study focuses on Kawagoe City, a typical medium-sized urban area in Saitama Prefecture, Japan. The traffic flow data used in this study were developed through detailed traffic simulations conducted as part of joint research with Kawagoe City. The dataset integrates road geometry, travel demand, and signal operations to represent realistic urban traffic conditions.

Intersection topology and lane configurations were obtained from OpenStreetMap, while origin–destination (OD) travel patterns (**Fig. 7(a)**) were estimated using a combination of building-use data and local traffic volume surveys conducted by Kawagoe City. Traffic assignment was performed using a user equilibrium model, run separately for each hour from 7:00 to 19:00 on both weekdays and weekends. This procedure generated 57,249 distinct routes, each associated with a flow volume.

To reflect realistic operating conditions at signalized intersections, signal patterns were calibrated based on observed traffic volumes and empirical signal timing ratios provided by the city. These data were used to construct probabilistic signal pattern distributions for each route, capturing the likelihood of stopping and queue-position-dependent dwell times. In addition, free-flow link speeds were adjusted to match observed travel times from the equilibrium assignment, thereby accounting for delays induced by traffic signals.

The resulting traffic flow distribution is shown in **Fig. 7(b)**. High traffic volumes are observed along the eastern bypass connecting to the highway interchange, while downtown streets—although also heavily used—are characterized by narrow widths and correspondingly lower operating speeds. **Fig. 8** presents the distributions of trip distances and energy consumption across all routes. The average trip distance is 2,967 m, and the average energy consumption is 307.3 Wh, values that are consistent with empirical measurements reported for urban EV operations.

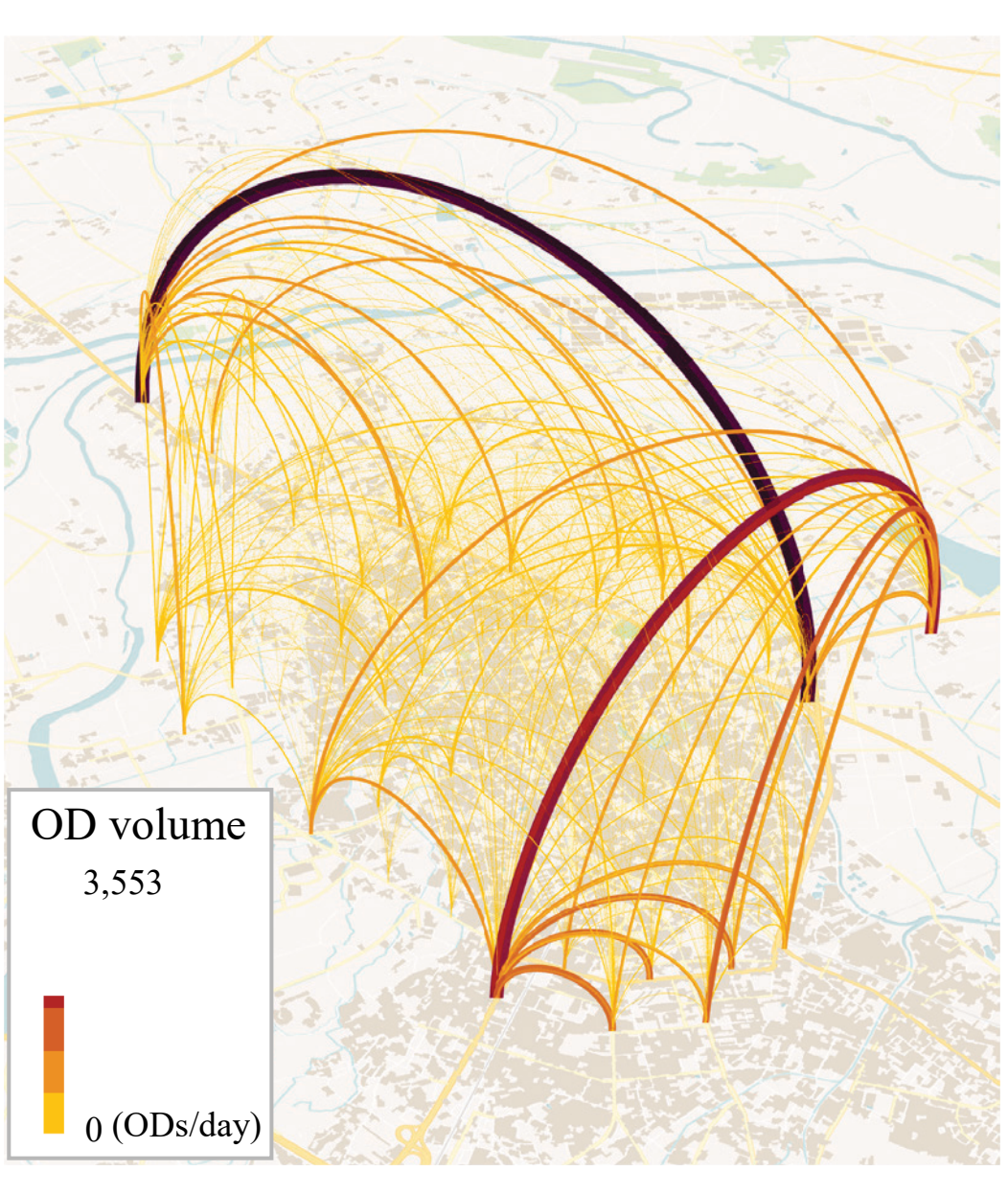

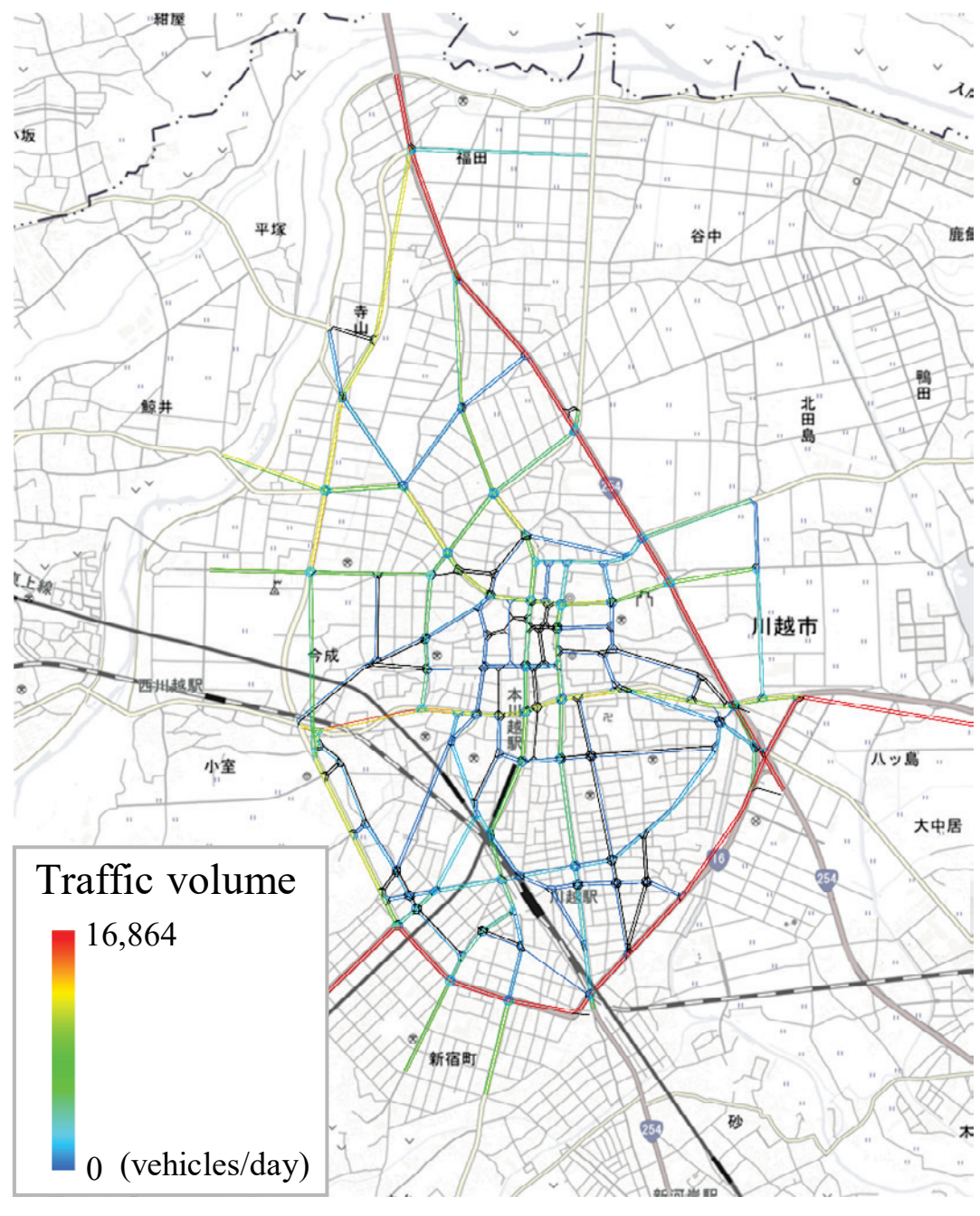

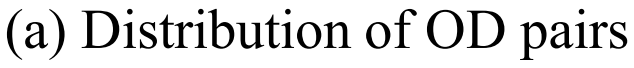
(a) Distribution of OD pairs

(b) Distribution of traffic volume

**Fig. 7. Traffic distribution in Kawagoe City**

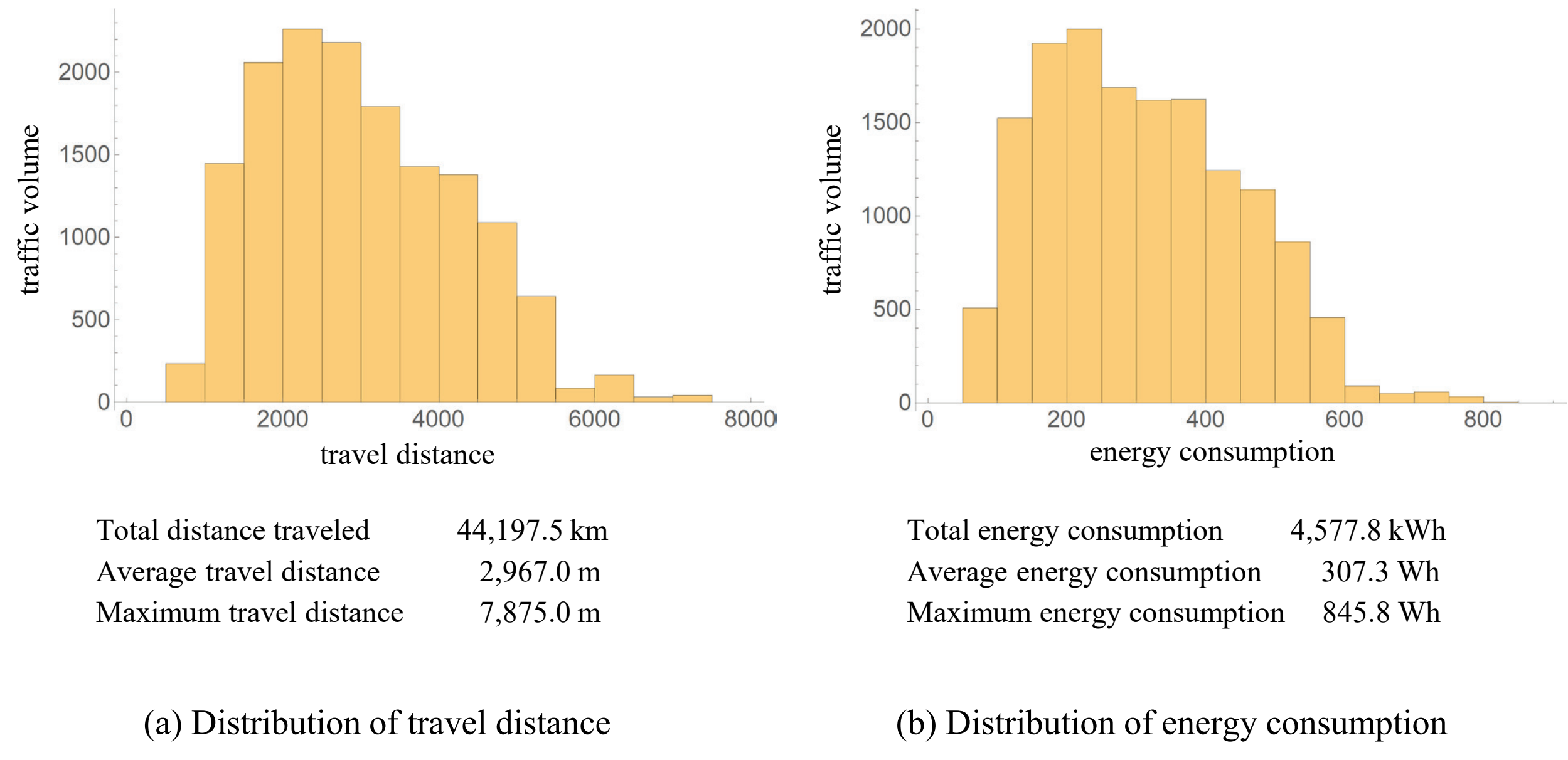

**Fig. 8. Distributions of travel distance and energy consumption**

## 5. Results

### 5.1. Problem size and computational environment

The Kawagoe road network covers a total length of 147,868 m. Discretizing the network into 7-m segments yields 21,124 binary decision variables $z_i$, each indicating whether a DWPT coil is installed on a candidate segment. All segments along the network are treated as candidate locations, including mid-block segments away from signalized intersections, so that the optimization itself confirms, rather than presupposes, where DWPT should be placed.

On the demand side, the dataset includes 57,249 routes obtained from hour-by-hour traffic assignment. Unless otherwise stated, these routes are aggregated into home-based groups (Section 3.3), which serve as the baseline definition of "groups" for the energy-balance constraints. Specifically, the Kawagoe OD data are generated at the road-link level: each link in the network acts as a virtual origin representing the residential population served by that segment, and all routes originating from a given link are grouped into one energy-balancing group. This link-level realization is finer than typical TAZ-based aggregation, while remaining coarser than individual building or household resolution. This combined scale is non-trivial for binary-location problems but remains tractable owing to the sparsity of the segment–route coupling, in which each variable $z_i$ appears in the energy balance only of routes traversing segment $i$. The model was solved using Gurobi 12.0.3 on a PC equipped with eight 2.2 GHz cores and 32 GB RAM. For each run, we imposed a computation time limit of 7,200 s. The reported solutions correspond to the best solutions found within the time limit. Across the nine scenarios reported in Section 5.3, the solver-reported optimality gaps remain small (0–2.47%), indicating that the solutions are near-optimal for the purpose of planning-level inference.

### 5.2. Baseline optimal location under home-based groups

**Fig. 9** shows the optimized DWPT location for the baseline setting (home-based groups; safety coefficient $\varphi = 1.1$). The solution confirms that signalized intersections are the most effective urban locations for DWPT, where deceleration, stopping, and queue discharge increase the time vehicles spend over embedded coils. Under this baseline case, the model equips 56 intersections with a total DWPT length of 2,233 m, corresponding to 1.51% of the total road network.

Consistent with Japan's left-hand traffic, many selected segments are placed on approach lanes that capture the dominant stopping and queuing behavior upstream of intersections. The deployment pattern is highly non-uniform across the city, reflecting heterogeneity in stop probability, queue formation, and signal timing. Even within the same corridor, the optimized solution allocates longer DWPT segments to approaches with higher red-encounter likelihood and longer expected dwell times, while assigning shorter segments where vehicles are more likely to pass during green phases. At Intersection A, for example, where a bypass road intersects with a major arterial, the installation length is relatively long due to the frequent occurrence of queues and extended stop durations. In contrast, other intersections along the same bypass, such as Intersection B, have shorter installations, reflecting longer green phases that reduce the likelihood of vehicles stopping. Downtown intersections, such as Intersection C, also receive DWPT installation because narrow street geometries, short block lengths, and recurrent congestion create conditions where vehicles spend more time idling and therefore present favorable opportunities for energy transfer.

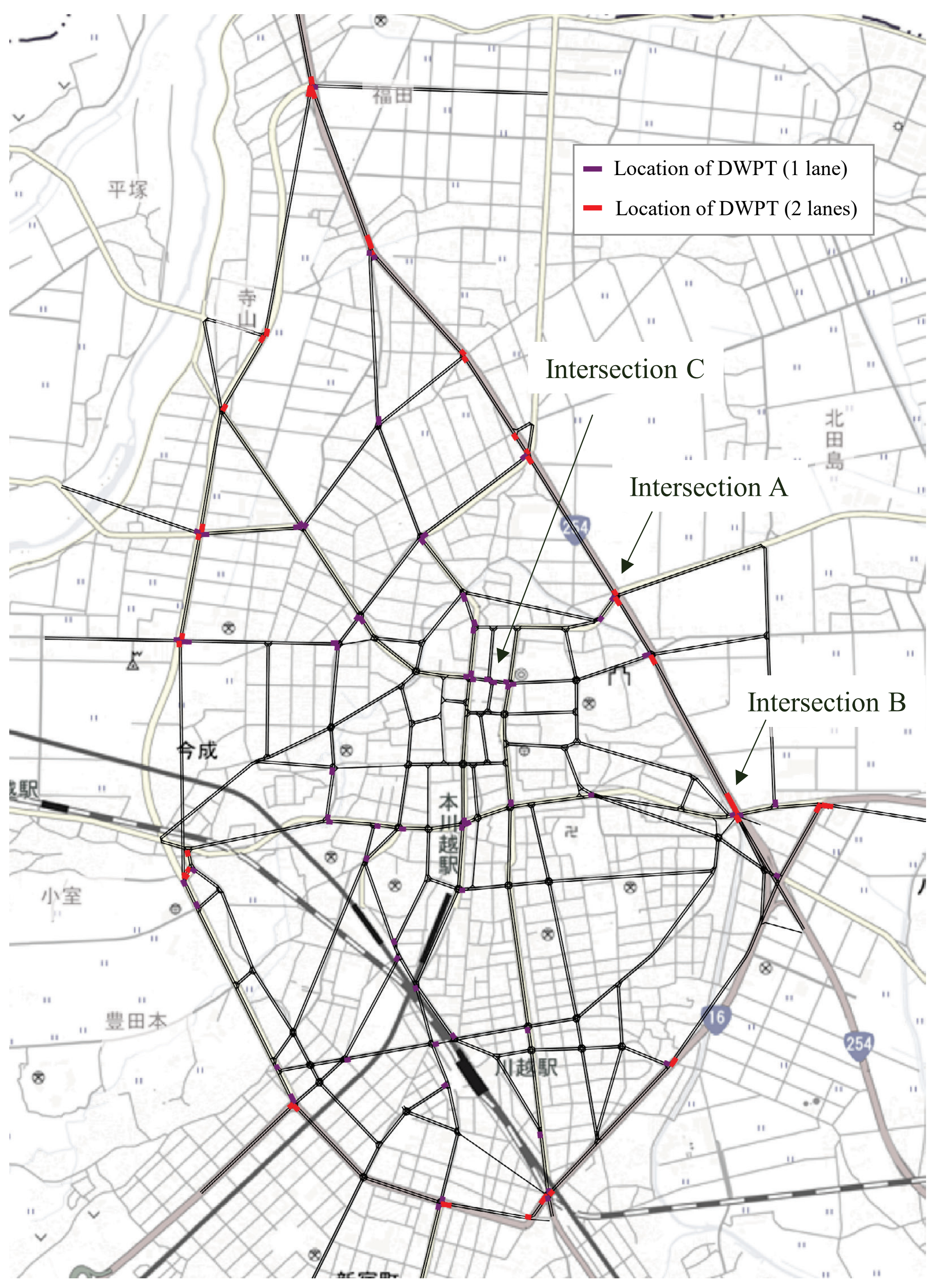

**Fig. 9. Optimal location of DWPT in Kawagoe City (home-based groups; safety coefficient $\varphi = 1.1$)**

### 5.3. Sensitivity to aggregation schemes and safety coefficients

To assess robustness to planning assumptions, we conduct a two-dimensional sensitivity analysis along (i) the aggregation scheme used in the energy-balance constraints and (ii) the safety coefficient $\varphi$. We consider three aggregation schemes: (a) entire region, (b) each home location (baseline), and (c) each OD pair, combined with three safety coefficient values $\varphi = 1.1, 1.3, 1.5$, resulting in nine scenarios.

**Table 3** summarizes the required DWPT length, the corresponding share of the road network, and the number of equipped intersections. Two consistent patterns emerge. First, for a fixed aggregation scheme, increasing $\varphi$ increases the required DWPT length. For example, under the baseline home-based aggregation, the required DWPT length rises from 2,233 m at $\varphi = 1.1$ to 3,346 m at $\varphi = 1.5$ (1.51% to 2.26% of total road length). Similar increases are observed for the region-based and OD-pair-based settings.

Second, for a fixed safety coefficient, more granular aggregation leads to larger infrastructure requirements. At $\varphi = 1.1$, the required DWPT length increases from 1,596 m (entire region) to 2,233 m (home-based) and 2,744 m (OD pair). At $\varphi = 1.5$, it increases from 2,387 m (entire region) to 3,346 m (home-based) and 4,291 m (OD pair). This reflects the intuitive fact that stricter, more granular energy-balance requirements reduce the degree to which surplus energy on some routes can offset deficits on others.

Importantly, even under the most demanding tested condition—OD-pair aggregation with $\varphi$ =1.5—the required DWPT length remains 4,291 m, corresponding to 2.90% of the urban road network. This indicates that the infrastructure requirement remains modest even when adopting conservative buffers and stringent aggregation assumptions.

**Table 3**
**Required DWPT length under different aggregation schemes and safety coefficients**

| Scheme | $\varphi$ | DWPT length (m) | Road share (%) | # Intersections | Optimality gap (%) |
|---|---|---|---|---|---|
| Region | 1.1 | 1,596 | 1.08 | 51 | 0.00 |
| Region | 1.3 | 1,981 | 1.34 | 55 | 0.00 |
| Region | 1.5 | 2,387 | 1.61 | 59 | 0.00 |
| Home (base) | 1.1 | 2,233 | 1.51 | 56 | 1.54 |
| Home (base) | 1.3 | 2,744 | 1.86 | 62 | 0.97 |
| Home (base) | 1.5 | 3,346 | 2.26 | 59 | 1.22 |
| OD pair | 1.1 | 2,744 | 1.86 | 63 | 2.47 |
| OD pair | 1.3 | 3,465 | 2.34 | 65 | 2.01 |
| OD pair | 1.5 | 4,291 | 2.90 | 68 | 1.89 |

(Note: $\varphi$ denotes the safety coefficient; DWPT length is the total installed length across the network.)

## 6. Operational validation and infrastructure–battery tradeoffs

The optimization model guarantees a non-negative expected energy balance at the group level but does not explicitly track the underlying battery state-of-charge (SOC) dynamics, since doing so would substantially increase computational complexity on the urban scale. With finite onboard battery capacity, however, short-term stochastic fluctuations may still lead to temporary depletion events even when the expected balance is satisfied. This section presents a simulation-based validation that addresses this gap by examining both the feasibility of the Infinite Drive concept under realistic SOC dynamics and the tradeoff between DWPT infrastructure investment and required onboard battery capacity.

### 6.1. Simulation design and evaluation metric

We perform Monte Carlo simulations of continuous urban operations by generating long trip chains, each comprising 100 round trips between a randomly selected home location and destinations drawn from the calibrated urban travel demand. These trip chains, with cumulative travel distance averaging approximately 600 km per run and reaching up to approximately 800 km, represent extended, high-utilization operations—such as taxi services, delivery fleets, ride-hailing vehicles, or future shared autonomous vehicles—that far exceed typical daily urban travel and therefore provide a conservative stress test for the Infinite Drive concept.

Each simulation run begins with the EV initially fully charged at its home location, after which no plug-in recharging occurs throughout the entire 100-round-trip sequence. Routes and departure times for each round trip are sampled from the calibrated demand, determining the signal-pattern probabilities along each path. Along each road segment, the SOC is updated by subtracting the segment-level energy consumption and adding energy transferred via DWPT when and where the segment is equipped and the corresponding signal pattern allows charging.

A run is defined as successful if the SOC never reaches zero throughout the entire 100-round-trip sequence. For each scenario, 1,000 independent simulation runs are conducted, and the success rate is calculated as the proportion of runs completed without battery depletion. Empirical success rates are reported together with 95% Clopper–Pearson exact confidence intervals (Clopper and Pearson, 1934), which remain well-defined at boundary cases such as 100% empirical success, where asymptotic approximations would degenerate.

### 6.2. Optimized versus uniform DWPT locations

We first compare the optimized DWPT location derived from the MIP model (home-based groups; safety coefficient $\varphi = 1.1$; total DWPT length = 2,233 m) with a simple uniform heuristic that installs 7 m of DWPT at every candidate signalized approach (total DWPT length = 3,584 m). Although the uniform heuristic requires substantially more DWPT infrastructure, it does not account for heterogeneity in queue formation, dwell-time distributions, or signal timing.

As shown in **Fig. 10**, for a 12 kWh battery, the optimized deployment achieves a success rate of 100.0% (95% CI: 99.63%–100.0%), whereas the uniform 7-m deployment achieves only 93.5% (91.79%–94.95%). For a 6 kWh battery, the optimized deployment achieves a success rate of 97.2% (95.98%–98.13%), compared to 84.1% (81.68%–86.31%) under the uniform deployment. The non-overlapping 95% confidence intervals between optimized and uniform deployments confirm that these differences are statistically significant under both battery configurations. Representative SOC trajectories are shown in **Fig. 10**, illustrating that failures under the uniform deployment occur more frequently and at earlier stages of continuous operation, despite the larger total DWPT length.

These results indicate that DWPT effectiveness depends not only on total installed length, but also on precise spatial alignment with signal-induced stopping behavior. Uniform deployment fails to exploit intersections where vehicles experience long dwell times and favorable queue positions, leading to missed charging opportunities even when more infrastructure is installed.

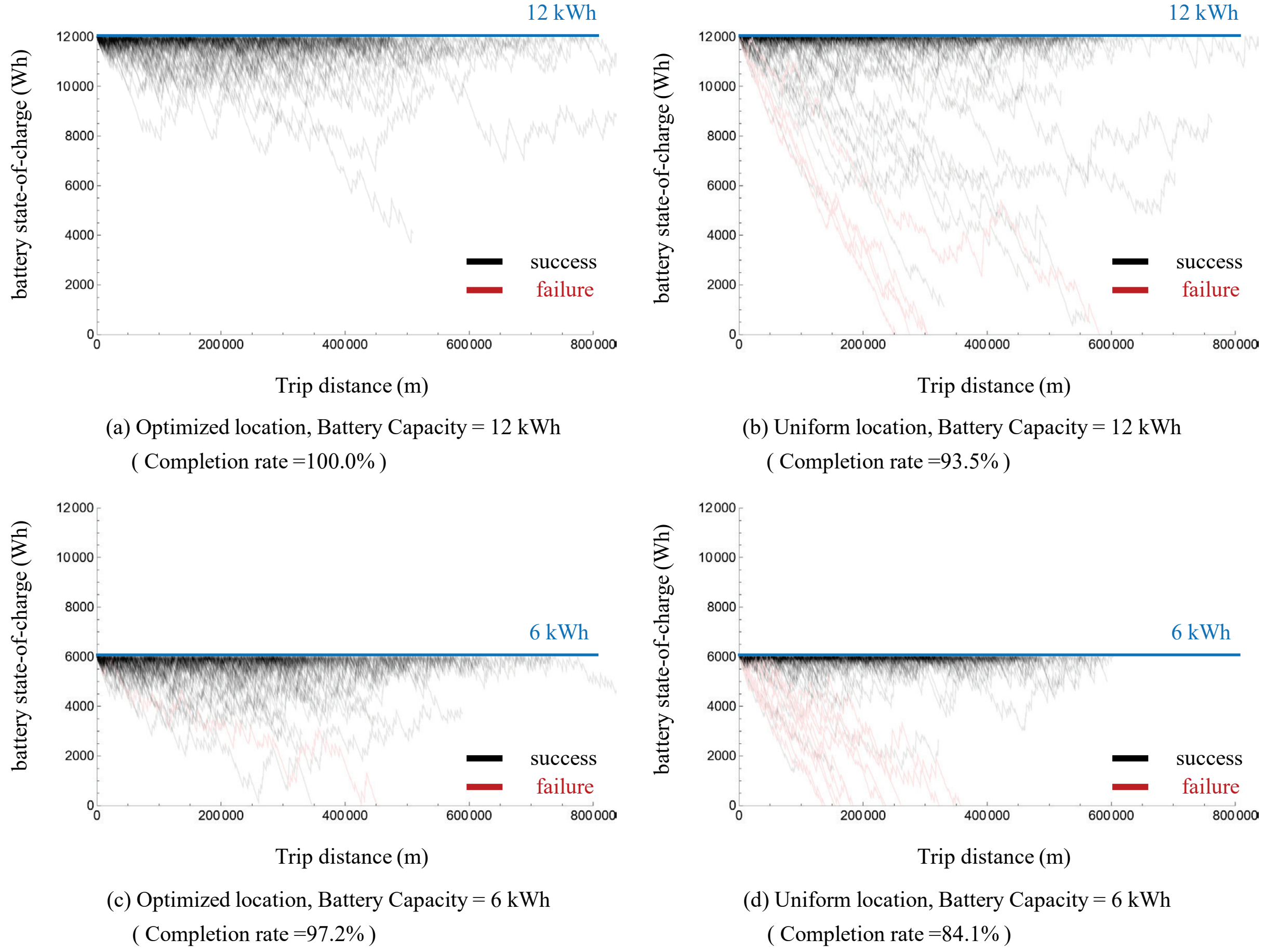

**Fig. 10. Comparison of battery state-of-charge (SOC) trajectories under optimized and uniform DWPT locations**

### 6.3. Infrastructure–battery tradeoffs under optimized deployment

We next examine the tradeoff between DWPT infrastructure length and EV battery capacity under optimized deployment strategies. **Fig. 11(a)** summarizes success rates as a function of battery capacity for optimized DWPT deployments under safety coefficients of 1.1, 1.3, and 1.5, corresponding to total DWPT lengths of 2,233 m, 2,744 m, and 3,346 m, respectively.

As the safety coefficient $\varphi$ increases, additional DWPT segments are installed to provide greater robustness against stochastic charging losses and finite battery constraints. This increase in infrastructure investment systematically improves reliability for a given battery size. Near-perfect performance can be achieved with progressively smaller batteries: approximately 12 kWh for $\varphi = 1.1$, 8 kWh for $\varphi = 1.3$, and 6 kWh for $\varphi = 1.5$. These results confirm that the safety coefficient in the optimization model provides a meaningful and interpretable mechanism for trading off infrastructure investment against vehicle-side energy storage.

For comparison, **Fig. 11(b)** presents analogous success-rate curves for uniform DWPT deployments of 7 m, 14 m, and 21 m per candidate approach. Achieving reliability comparable to the optimized solutions requires substantially larger DWPT investments—up to more than 7.3% of the total road length—highlighting the inefficiency of uniform location strategies. Importantly, the observed tradeoff between DWPT length and battery capacity represents an environmental outcome in itself. By enabling reliable operations with substantially smaller onboard batteries and a limited electrified road footprint, the optimized solutions demonstrate system-level potential for reducing vehicle mass, material demand, and infrastructure-related impacts.

To illustrate the potential scale of the vehicle-side environmental co-benefit, we draw on the life-cycle assessment literature on Li-ion EV battery manufacturing. Recent cradle-to-gate (i.e., manufacturing-stage) emission factors for commercial-scale NMC-chemistry Li-ion batteries are estimated at 61–106 kg $CO_2$e/kWh depending on the electricity mix used in cell production (Emilsson and Dahllöf, 2019); earlier estimates of 140–172 kg $CO_2$e/kWh reflected less efficient cell manufacturing and higher-carbon electricity inputs (Ellingsen et al., 2014; Kim et al., 2016). Using the most recent range, the 12 kWh battery sufficient for Infinite Drive operation at $\varphi = 1.1$ corresponds to approximately 1.7–3.0 tonnes $CO_2$e of avoided battery manufacturing per vehicle relative to a conventional 40 kWh urban EV. The magnitude of this vehicle-side benefit is broadly consistent with the DWPT life-cycle assessment of Bi et al. (2019), who showed that selective electrification of roughly 3% of road lane-miles can enable substantial battery downsizing while reducing fleet-wide greenhouse gas emissions. This estimate does not include the embodied emissions of the DWPT road infrastructure itself, which remain poorly characterized in the literature and lie beyond the scope of this study; a full comparative life-cycle assessment is therefore left for future work. The figure should be interpreted as an upper bound on the vehicle-side co-benefit rather than as a net environmental accounting. Joint optimization of charger location and battery size in electric-bus systems has also been studied as a precursor to the present co-design framing (Liu et al., 2017), while Liu et al. (2024) recently extended this co-design perspective to passenger EVs by jointly configuring DWPT facilities and battery capacity in a road-network setting. Comparable life-cycle analyses of DWPT in electric bus transit systems have similarly highlighted that infrastructure–battery tradeoffs and environmental performance depend critically on operational context, including service frequency and infrastructure sharing across multiple lines (Pei et al., 2024).

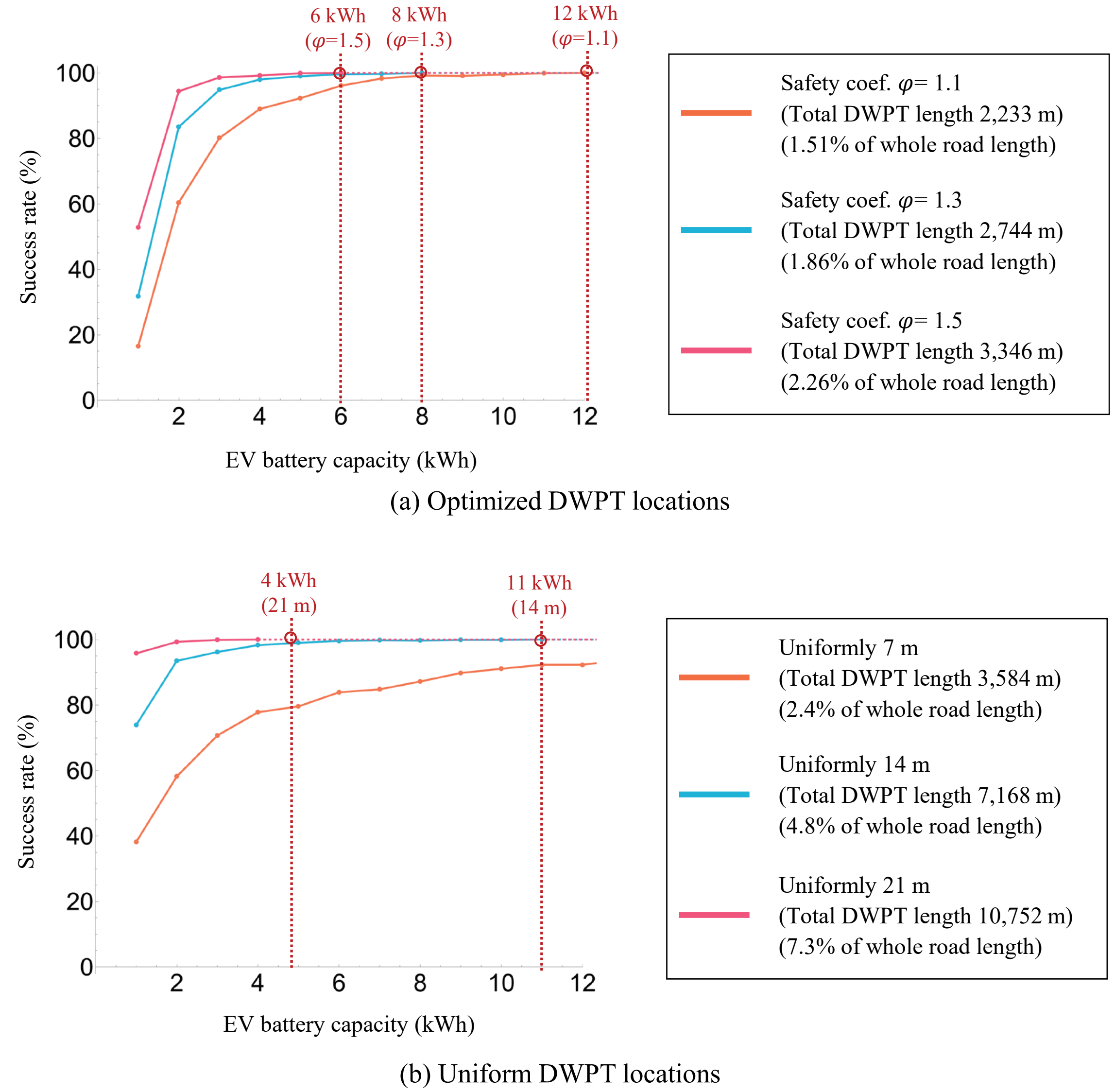

**Fig. 11. Tradeoff between DWPT infrastructure length and required battery capacity**

### 6.4. Implications for battery sizing and infrastructure planning

Taken together, these results provide an operational validation of the Infinite Drive concept and demonstrate that the proposed framework can be used not only to minimize DWPT length, but also to inform joint decisions on infrastructure deployment and battery sizing. Rather than treating battery capacity as an exogenous design choice, planners can use the framework to evaluate how targeted DWPT deployment enables smaller batteries without compromising operational reliability.

From a sustainability perspective, this tradeoff has important implications. Smaller onboard batteries imply lower vehicle mass, reduced material demand, and potentially lower upstream manufacturing impacts, while optimized DWPT deployment minimizes the physical footprint and embodied impacts of road electrification. In fact, the cost savings from reducing battery capacity can help pay for the DWPT infrastructure. By shifting part of the energy sufficiency burden from vehicle-side overdesign to system-level infrastructure planning, optimized DWPT deployment offers a pathway toward more cost-effective, resource-efficient, and environmentally sustainable urban EV systems.

## 7. Discussion

### 7.1. Comparison with prior DWPT location work

The methodology and findings presented in this study can be situated relative to several streams of prior DWPT location research. Studies of DWPT deployment at metropolitan and corridor scales (Fuller, 2016; Honma et al., 2024; Mubarak et al., 2021; Trinko et al., 2022; Yan et al., 2022) have established the feasibility and economic rationale of large-scale DWPT deployment, but typically operate at coarser spatial resolution that does not explicitly capture intersection-driven stop-and-go dynamics. The present framework complements these by introducing intersection-level resolution through 7 m segments combined with probabilistic signal patterns and saturation headway-based queue modeling, capturing both queue-position-dependent dwell times and signal-induced acceleration/deceleration that shape charging opportunities in dense urban networks.

A parallel stream of equilibrium-based urban DWPT location models (Chen et al., 2016; Liu et al., 2021; Riemann et al., 2015; Tran et al., 2022) endogenizes driver route choice in response to deployment. The present framework occupies a complementary methodological position: by accepting fixed route flows as exogenous input, it can devote computational resources to substantially finer spatial resolution, while remaining naturally extensible to equilibrium-based assignment once DWPT deployment density grows high enough to materially alter routing behavior.

A more recent stream emphasizes joint optimization of DWPT infrastructure and battery capacity, co-determining charger locations and battery sizes for electric-bus systems (Li et al., 2024; Liu et al., 2017) and, most recently, for passenger EVs operating on road networks (Liu et al., 2024). The infrastructure–battery tradeoff demonstrated through the SOC simulations in Section 6 contributes to this line of work by adding Monte Carlo–based operational validation of long trip chains, rather than relying solely on expected-value optimization guarantees. Taken together, three features collectively distinguish the present contribution from prior streams: (i) the introduction of "Infinite Drive" as a complete-coverage design principle for urban EV infrastructure; (ii) explicit modeling of signalized intersection dynamics through probabilistic signal patterns and saturation headway-based queue representation; and (iii) operational validation through SOC simulations that complements expected-value optimization with finite-battery feasibility analysis.

### 7.2. Generalizability beyond Kawagoe

The 1.51% baseline result obtained for Kawagoe reflects the characteristics of a medium-sized Japanese city with dense signalized intersections and Japan-typical OD demand patterns. The qualitative finding—that targeted DWPT location at signalized intersections enables Infinite Drive with minimal infrastructure—is expected to hold across urban contexts, but the specific share of the road network requiring electrification will depend on local intersection density, arterial-versus-grid network topology, and trip-distance distributions. Cities with sparser intersection density may require relatively more DWPT coverage because the queue-position-dependent dwell-time mechanism that drives our efficiency results is most effective at signal-rich nodes. Conversely, cities with shorter typical trips may require less coverage. The framework itself is directly applicable to any signalized urban network with comparable input data, and future cross-city comparative studies would help establish quantitative scaling relationships between urban form characteristics and required DWPT share.

### 7.3. Limitations and assumptions revisited

Several assumptions underlying our framework warrant explicit acknowledgment. First, routes and traffic flows are treated as exogenous (Section 3.1), which is reasonable for low-to-moderate DWPT density but becomes a first-order limitation if deployment is dense enough to materially alter driver routing decisions. Second, the signal-pattern probabilities assume uniform vehicle arrival distributions within each cycle (Section 3.4), an appropriate approximation for the Kawagoe context but one that would overestimate back-of-queue dwell times in coordinated-arterial networks where platoon arrivals predominate. Third, the vehicle parameters in Table 2 represent a passenger sedan; freight, transit, and other multi-class fleets would require parameter sweeps falling outside this study's single-class focus. Fourth, the operational validation in Section 6 employs a single home-based trip-chain structure consisting of 100 round trips per run; fleet operations involving depot-based dispatch or shared autonomous vehicles would warrant separately tailored validation scenarios. These limitations define natural directions for the future research discussed below.

### 7.4. Future research directions

**Methodological extensions**

Building on the limitations identified above, five methodological directions merit future investigation: (1) integration with dynamic traffic assignment to capture time-varying congestion effects (Wang et al., 2018); (2) application of multi-layer tensor architectures (Zhou et al., 2025) for joint optimization of DWPT locations, grid connections, and routing; (3) chance-constrained formulations to address battery capacity limitations more rigorously than the current safety coefficient approach; (4) computational scaling to larger metropolitan networks by restricting candidate segments to intersection-vicinity locations; and (5) multi-class extension to heterogeneous fleets, requiring integration of vehicle-class-specific energy parameters with heterogeneous queue and segment representations to capture vehicle-length effects on the 7-m discretization, queue-position counting, and saturation headway.

**Environmental assessment extensions**

The segment-level kinematics developed in this study provide direct inputs for environmental impact quantification. Following the cross-resolution approach (Zhou et al., 2015), future work can couple these trajectory outputs with emission rate functions to assess mixed-fleet impacts during the EV transition period. The stochastic parameters introduced by Meng et al. (2021) could extend our signal pattern formulation to capture driver variability. Additionally, a full comparative life-cycle assessment—extending the preliminary battery-side estimate presented in Section 6.3 to include DWPT road infrastructure, installation, and use-phase emissions—would quantify the net embodied carbon and material benefits. Such an assessment would also support a marginal-abatement-cost analysis of urban DWPT deployment in policy-relevant cost-per-tonne $CO_2e$ terms, connecting the framework to the climate-policy literature (IPCC, 2022). Importantly, the operational validation presented in Section 6 demonstrates that such battery downsizing scenarios are not merely theoretical, but operationally feasible under realistic urban travel conditions.

**Economic considerations**

Although a detailed cost analysis is beyond the scope of this study, the infrastructure–battery tradeoff demonstrated here carries direct economic implications. Smaller onboard batteries reduce vehicle purchase costs, with current Li-ion battery pack prices on the order of $100–150 per kWh (BloombergNEF, 2024; Ziegler and Trancik, 2021) implying potential per-vehicle savings of several thousand US dollars when battery capacity is reduced from 40 kWh to 12 kWh. These savings must be weighed against DWPT deployment costs, which include infrastructure installation, maintenance, and—importantly—repair costs, since damaged or malfunctioning coils embedded in road surfaces can be more difficult and costly to access than stationary chargers. Conversely, DWPT eliminates the downtime associated with plug-in charging, which can be substantial for high-utilization fleets and represents a significant economic value beyond direct energy delivery. A full system-level cost-benefit analysis integrating vehicle-side savings, DWPT capital and operational expenditures, and downtime valuation is left as a direction for future work, drawing on existing economic feasibility studies (Honma et al., 2024; Trinko et al., 2022).

## 8. Conclusions and policy implications

This study developed a new MIP model for optimally locating DWPT infrastructure in urban areas, with the aim of realizing the concept of Infinite Drive—a scenario in which EVs can complete continuous urban operations without ever requiring plug-in charging. Application to Kawagoe City, Japan, demonstrated that Infinite Drive can be achieved with a remarkably modest amount of infrastructure. Under the baseline home-based aggregation and a safety coefficient of $\varphi = 1.1$, only 2,233 m of DWPT, or 1.51% of the road network, distributed across 56 intersections was sufficient to sustain all EV operations in expectation; even under the most demanding tested condition—OD-pair aggregation with $\varphi = 1.5$—the required installation length remained 4,291 m, corresponding to 2.90% of the total road network. Monte Carlo simulations of continuous trip chains further confirmed that the optimized deployment can sustain long-duration operations with finite batteries, with near-perfect performance achievable with approximately 12 kWh, 8 kWh, and 6 kWh batteries when total DWPT length is 2,233 m, 2,744 m, and 3,346 m, respectively.

Beyond demonstrating technical feasibility, the findings highlight the practical relevance of DWPT as a component of sustainable urban mobility systems. The benefits are especially pronounced for high-utilization vehicles, such as taxis, delivery fleets, ride-hailing services, and future autonomous vehicles (AVs), which operate continuously and often lack convenient access to fixed charging infrastructure. For these users, Infinite Drive provides independence from plug-in stations, increasing vehicle productivity and asset utilization.

The concept also carries important environmental and policy implications. Continuous in-motion charging reduces the need for oversized batteries, lowering vehicle weight, improving energy efficiency, and reducing reliance on scarce materials. As illustrated in Section 6.3, the corresponding vehicle-side reduction in battery manufacturing emissions is on the order of 1.7–3.0 tonnes $CO_2e$ per vehicle relative to a conventional 40 kWh urban EV, representing a front-loaded co-benefit that does not depend on future grid decarbonization. It enables more direct use of surplus renewable energy, particularly midday solar generation, while reducing idling and queuing at stationary charging facilities. While the realized magnitude of this synergy varies by local climate and renewable-energy mix, the continuous nature of DWPT provides the temporal flexibility needed to align EV demand with whichever renewable generation pattern characterizes a given urban context. As a result, DWPT can help integrate EVs more efficiently into urban energy systems and align transport policy with broader goals of carbon reduction and sustainable urban mobility.

For city planners considering DWPT deployment, three concrete recommendations emerge from this analysis. First, DWPT segments should be located at signalized intersections with frequent red-light encounters and substantial queue formation, rather than uniformly distributed across all approaches; as shown in Section 6.2, uniform deployment requires substantially more infrastructure while delivering lower operational reliability. Second, the choice of aggregation scheme in the optimization should be matched to the dominant vehicle class served: anchor-based aggregation for home- or depot-anchored fleets, OD-pair aggregation for recurring commute patterns, and region-based aggregation for shared mobility and autonomous services. Third, DWPT deployment decisions should be co-designed with target battery capacity rather than treating the two as independent design variables, since modest increases in installed DWPT length can enable substantial reductions in required onboard battery capacity, yielding combined cost and environmental benefits. Taken together, these findings position targeted DWPT deployment not as supplemental charging infrastructure, but as a strategic enabler of plug-free, low-carbon, and resource-efficient urban mobility systems.

**Acknowledgments**

This work was supported by the Japan Society for the Promotion of Science (JSPS), Grant-in-Aid for Scientific Research (B) 24K01109.